\documentclass[]{aa} 
\usepackage{graphicx}
\usepackage{txfonts}
\usepackage[colorlinks,citecolor=blue,urlcolor=blue]{hyperref}
\usepackage{verbatim}
\usepackage{float}
\newcommand{\kms}{\,km\,s$^{-1}$\xspace} % km/s
\newcommand{\Zsun}{$\,Z_\odot$\xspace}  % solar metallicity
\newcommand{\bse}{{\tt BSE}}

\newcommand{\ssebse}{{\tt SSE/BSE}}
\def\abhtwo{a_{\rm BH,2}}
\def\abhone{a_{\rm BH,1}}
\def\abh{a_{\rm BH,i}}

\newcommand{\Ms}{\ensuremath{{\rm M}_{\odot}}}
\def\qcrit{q_{\rm crit}}
\def\etahe{\eta_{\rm HeBH}}
\def\nhe{N_{\rm HeBH}}
\def\nphe{N^{\prime}_{\rm HeBH}}
\def\nheobs{N_{\rm HeBH,obs}}
\def\sfr{{\mathcal R}_{\ast,0}}
\def\tauhe{\tau_{\rm He}}
\newcommand{\peryr}{\ensuremath{{\rm~yr}^{-1}}}

\begin{document}

   \title{X-ray emission from helium star+black hole binaries as probes of tidally induced spin-up of second-born black holes}
   \titlerunning{X-ray emission from He+BH binaries}
   \author{K. Sen
          \inst{1,2}
          \and
          A. Olejak
          \inst{3}
          \and
          S. Banerjee
          \inst{4,5}
          }

   \institute{Institute of Astronomy, Faculty of Physics, Astronomy and Informatics, Nicolaus Copernicus University, Grudziadzka 5, 87-100 Torun, Poland
         \and
             Steward Observatory, Department of Astronomy, University of Arizona, 933 N. Cherry Ave., Tucson, AZ 85721, USA \\
                \email{ksen@arizona.edu}
        \and 
             Max Planck Institute for Astrophysics, Karl-Schwarzschild-Strasse 1, 85748 Garching, Germany
        \and
             Helmholtz-Instituts f\"{u}r Strahlen- und Kernphysik, Nussallee 14-16, D-53115 Bonn, Germany
        \and
             Argelander-Institut f\"{u}r Astronomie, Auf dem Hügel 71, D-53121, Bonn, Germany
             }

   \date{Received \today; accepted ...}

% \abstract{}{}{}{}{} 
% 5 {} token are mandatory
 
  \abstract
   {Tidally induced spin-up of stripped helium stars in short-period (<\,1\,d) binaries with black holes (BHs) has been presented as one of the possible mechanisms to reproduce the high-spin tail of the BH spin distribution derived from gravitational wave (GW) merger observations. At such short periods, a fraction of the strong stellar wind from the stripped helium stars may be accreted by the BHs, and its gravitational potential energy may be released as observable radiation in the X-ray regime. }
   {We estimate the X-ray luminosity and its observability from the population of BHs in orbit with stripped helium stars that evolve into binary BH or BH+neutron star binaries and merge within a Hubble time. }
   {We post-process recent advancements for estimating X-ray luminosities (via wind accretion onto stellar mass BHs) into the rapid population synthesis codes BSE and StarTrack. We derive lower limits on the X-ray luminosity distribution from the above population of stripped helium star+BH binaries at four metallicities (0.01, 0.1, 0.5, 1 $Z_{\odot}$) and two mass transfer stability criteria. }
   {We find that a large fraction (0.1-0.5) of stripped-helium stars in the above population transfer enough wind matter onto the BH to produce X-ray luminosities above $10^{35}$\,erg\,s$^{-1}$, up to $\sim10^{39}$\,erg\,s$^{-1}$. Such binaries should be observable as X-ray bright systems at 0.1\,$Z_{\odot}$, 0.5\,$Z_{\odot}$ and $Z_{\odot}$, representative of Sextans A, the Large Magellanic Cloud (LMC) and the Solar neighbourhood, respectively. We show that most of these X-ray-bright systems also have the shortest orbital periods where tides can spin up the stripped helium star component. The formation efficiency of these systems increases with decreasing metallicity. However, accounting for the local star formation rates, our population synthesis predicts $\sim$2 and $\sim$1 such X-ray-bright helium star+BH binaries in the Milky Way and LMC, respectively, that will produce a binary compact object merger within a Hubble time. } 
   {Ongoing high-sensitivity X-ray surveys and high-resolution optical surveys of low metallicity environments such as Sextans A is an important stepping stone to identifying the population of short-period helium star+black hole binaries and possibly constrain the contribution of isolated binary evolution to the high spin tail of the black hole spin distribution in GW mergers. }


   \maketitle
%
%-------------------------------------------------------------------

\section{Introduction}

The observed spin of black holes (BHs) is a rapidly evolving field of study 
\citep[e.g.][]{Reynolds2021,Genzel2024}. In particular, there is an apparent dichotomy 
between the spins of stellar-mass BHs found 
in X-ray binaries and gravitational wave mergers \citep[][]{Orosz2009,
Orosz2011,abbott2016,abbott2019,Roulet2021,Miller-Jones2021,abbott2023,Draghis2024}- 
the BHs in X-ray binaries are estimated to be more rapidly 
spinning than those in gravitational wave mergers, \citep[see however][]
{Zdziarski2024,Belczynski2024}. While BHs in low-mass X-ray binaries can 
be spun up by accretion \citep{podsiadlowski2003,Oshaughnessy2005,Fragos2015}, 
theoretical studies imply that BHs in high-mass X-ray binaries and gravitational 
wave mergers may form distinct populations and/or there exist observational 
selection effects \citep{Fishbach2022,Liotine2023,Misra2023,Romero2023}. 
As such, these systems are an excellent tested for stellar evolution theory 
\citep[for a recent review, see][and references therein]{Mandel2024}. 

While most of the gravitational wave merger events have been detected with 
low effective spin $\chi_{\rm eff}$\footnote{defined as \citep{Racine2008,Santamaria2010,Ajith2011}
\begin{equation}
    \chi_{\rm eff} = \frac{m_{\rm 1}\Vec{\chi}_{\rm 1} + m_{\rm 2}\Vec{\chi}_{\rm 2}}{m_{\rm 1}+m_{\rm 2}} \cdot \hat{L}
\end{equation}
where $m_{\rm 1}$, $m_{\rm 2}$, $\Vec{\chi}_{\rm 1}$ and $\Vec{\chi}_{\rm 2}$ are the masses and spin vectors of the component black holes and $\hat{L}$ is the unit vector of the Newtonian angular momentum of the binary.
}
\citep[$\chi_{\rm eff}\,\sim$\,0.05,][]{Roulet2019,abbott2023} and thus 
provide evidence for efficient angular momentum coupling between the core 
and envelope of the BH progenitors (\citealp{Spruit2002,Fuller2019}, see 
also \citealp{Skoutnev2024a,Skoutnev2024b}), there exists a tail 
in the spin distribution, reaching values that are up to unity. The origin of this 
long tail is not solved and may yet still be model dependent \citep[e.g., see][]{
Zevin2021,Callister2022}. Recent literature suggests that a high-spin tail 
could imply a contribution from the isolated binary evolution channel with tidally 
spun-up second born BH \citep{Bavera2020,Olejak2021b} or dynamically induced multiple 
mergers e.g. in globular clusters or AGN disks \citep{Mapelli2020,Banerjee2021,
Mandel2022,Santini2023,Kiroglu2024,Delfavero2024,Li2025}. An asymmetric distribution 
of effective spins \citep{Roulet2021} implies that a substantial contribution may 
originate from isolated binary evolution. 

In the isolated binary evolution channel, binary black holes (BBHs) that merge 
within a Hubble time can form via one or more phases of stable mass transfer 
\citep[SMT,][]{Heuvel2017,Gallegos-Garcia2021,Bavera2021,Olejak2021a,VanSon2022,Shao2022,Briel2023,
Picco2024,Dorozsmai2024} or involves a common envelope phase \citep[CE,][]{Tutukov1993,Giacobbo2018,
kruckow2018,Spera2019,Ginat2020,Belczynski2020,Renzo2021,Gallegos-Garcia2023,
Boesky2024,Romagnolo2024}. In both of the above subchannels, the BBH phase is typically preceded 
by a helium star+BH (He+BH) phase where the He+BH binary has an orbital period 
less than $\sim$1\,d \citep{Korb2024}. It has been suggested that He stars in 
short period binaries can be spun up by tides, such that the Kerr 
parameter of the second-formed BH can reach close to unity \citep{Detmers2008,
Kushnir2016,Qin2018,Belczynski2020,Bavera_2021b,Olejak2021b,Fuller2022,Ma2023}. 

Although the stellar wind from stripped helium stars is expected to be very 
fast \citep[e.g., see][]{Sander2012}, the short periods imply that the Bondi 
accretion radius of the BH \citep{Bondi1944,Shakura1973} can subtend a 
solid angle large enough to capture a substantial amount of wind mass lost 
from the He stars. As such, observable X-ray emission can be produced from 
the shortest period He+BH binaries that would also potentially contribute to 
the tail of the BH spin distribution in gravitational wave merger data due 
to efficient tidal spin-up of the second-formed BH’s progenitor He star. 
Previous studies have largely focused on X-ray emission from He+neutron star 
binaries \citep{Shao2015,Shao2019ULXs,Mondal2020,Chen2023,Misra2024,Li2024}, 
BHs with nondegenerate star companions \citep{vanbeveren2020,Sen2021,Shao2019BHB,
Liotine2023,Misra2023,Romero2023,Kruckow2024,Xing2025}, or BH Ultra Compact 
Binaries, where X-rays are produced from mass accretion due to Roche Lobe 
Overflow \citep{Shao2020,Wang2021,Qin2024}.

Recent studies have shown that observable X-ray emission can be produced 
from stellar-mass BHs even at low mass accretion rates \citep{Xie2012,Sen2024}. 
In this work, we aim to study the production and observability of X-ray 
emission from short-period He+BH binaries and its correlation with the tidal 
spin-up of the He stars and the spins of their remnant BHs. Our 
work is particularly relevant for ongoing \textit{Chandra} surveys of the 
Large Magellanic Cloud ($\sim Z_{\odot}/2$, \citealp{Choudhury2021}) and 
Sextans\,A ($\sim Z_{\odot}/10$, \citealp{Kaufer2004,Lorenzo2024}), which is slated to 
detect X-ray point-sources down to $\sim$10$^{32}$\,erg\,s$^{-1}$ \citep{Antoniou2022} 
and $\sim$10$^{35}$\,erg\,s$^{-1}$ \citep{Antoniou2023}, respectively. 
Soon, proposed missions such as AXIS \citep{Mushotzky2018}, HEX-P 
\citep{Madsen2018,Lehmer2023}, or STROBE-X \citep{Ray2019} may add further 
observational constraints for our model predictions. 

The paper is organised as follows. Section\,\ref{section_method} discusses 
the rapid binary evolution models and the X-ray prescription used to model 
the He+BH phase. We present the distribution of X-ray luminosities and 
predictions from our population synthesis calculations in Sect.\,\ref{section_result}. 
We give our take-home messages in Sect.\,\ref{section_conclusion}. 

%--------------------------------------------------------------------

\section{Methods}
\label{section_method}

We synthesize populations of BBH mergers and their He+BH progenitors utilising 
the fast binary evolution codes $\bse$ \citep{Hurley_2000,hurley2002} and 
StarTrack \citep{Belczynski2008,Belczynski2020}. We choose the initial 
stellar binary distribution according to \citet{sana2012}, namely, orbital 
periods within 0.15\,$\leq$\,$\log_{10}(P/{\rm d})$\,$\leq$\,3.5 %\footnote{in StarTrack this range is extended to 5.5 as proposed by \cite{deMink2015}.} 
following the distribution $f(\log_{10}P) \propto (\log_{10}P/{\rm d})^{-0.55}$ and 
eccentricities following $f(e) \propto e^{-0.45}$. The initial binary components 
are zero-age main sequence (ZAMS) stars, distributed according 
to the Salpeter mass function \citep{salpeter1955}, $f(m) \propto (m/\Ms)^{-2.35}$, 
over the range 5\,$\Ms$ to 150\,$\Ms$, %(primary) and 5\,$\Ms \leq m_2 \leq$\,150\,$\Ms$ (secondary), 
and are paired with each other randomly. We evolve $10^6$ binaries for each 
of the four metallicity 
values $Z=$\Zsun, 0.5\Zsun, 0.1\Zsun, and 0.01\Zsun (here,\Zsun = 0.02), corresponding to the 
total stellar mass (integrated over all masses) of $\sim9.6\times10^{7}$\,M$_{\odot}$ 
(for a binarity fraction of 100\% ). 

\subsection{BSE models}
\label{bse}

We use an updated version of $\bse$, as described in \citet{Banerjee2024}, 
derived from the original version of $\ssebse$ \citep{Hurley_2000,hurley2002}. 
The updated $\bse$ contains several extensions regarding stellar wind, 
stellar remnant mass, binary mass transfer physics, and the natal spin of BHs, 
described in detail in \citet{Banerjee2024}.

During a mass transfer episode, the mass transfer and accretion rates
are limited by the thermal timescales of the donor and
the accretor or the dynamical timescale \citep{hurley2002}. Apart from this limit,
$\bse$ assumes a $\beta=100$\% mass accretion efficiency onto 
a non-degenerate accretor. As for accretion onto a degenerate or
compact member (a WD, NS, or BH), the mass accretion rate is capped 
by the Eddington limit. Any mass that is not accreted onto the recipient
is assumed to be lost from the system altogether, and this mass loss
extracts orbital angular momentum from the system. For mass transfer
onto a non-degenerate companion, we assume that the lost
material carries with it the specific angular momentum of the donor,
as defaulted in $\bse$ (the `$\gamma=-1$' option). For super-Eddington-rate mass transfer
onto a degenerate or compact companion, the released material carries the specific angular momentum of the accretor (as if it is a wind
from the accretor). 
%(In the updated $\bse$ described in \citealt{Banerjee2024},
%it is possible to reduce $\beta$ to obtain a more non-conservative
%mass transfer. However, in the present $\bse$ runs, $\beta=100$\%
%is always used.)

The birth masses of NS and BH
are derived according to the `delayed' remnant mass model of \citet{Fryer2012},
incorporating the pair-instability and pulsation pair-instability supernova
models of \citet{Belczynski_2016a}. The remnant masses are also influenced
by the stellar wind mass loss that incorporates, among other ingredients (see
\citealt{Hurley_2000,Banerjee_2020}), the \citet{Vink2001} bi-stability-jump
wind model for O-type stars and the wind Eddington factor of \citet{Graefener_2011} 
(as implemented in \citealp{Giacobbo2018}). The remnant natal kicks
are drawn from a Maxwellian velocity distribution of $\sigma=265$\kms \citep{Hobbs2005},
which are then reduced based on the SN fallback fraction \citep{Fryer2012}.

In these calculations, BHs are, in general, born with a low spin magnitude
(Kerr parameter, $\abh$ = |$\Vec{\chi}_{\rm i}$|; $i=1$ or 2) as obtained 
by the fast-rotating MESA
massive-star models of \citet{Belczynski2020}. Here, owing to efficient
angular momentum transport from the stellar core driven by the Tayler-Spruit 
dynamo \citep{Spruit2002}, BHs are born with $0.05\lesssim\abh\lesssim0.15$.
However, when a `second-born' BH is formed from a tidally spun-up He star 
in a close He+BH binary (orbital period $<1.0$ day), the BH is born with a 
potentially higher spin, $\abhtwo$\,$\leq$\,$1.0$, based on the BH spin-up 
model of \citet{Bavera2020,Bavera_2021b}.

The default model of the maximum donor-to-recipient mass ratio for a binary to undergo
stable mass transfer, $\qcrit$ \citep{hurley2002}, produces BBH mergers 
predominantly ($>90\%$) via the CE channel (see, e.g., \citealt{Giacobbo2018,
Olejak_2020,Banerjee_2021b}). In one set of binary-evolutionary models, we 
adopt this default $\qcrit$ model (that allows Hertzsprung-gap donors to 
undergo a CE) and designate the resulting BBH-merger population and its 
progenitor population as `CE-dominated'. In a second set of population 
synthesis runs, we take a constant, high $\qcrit=8$, as in \citet{Banerjee2024}, 
that instead results in BBH mergers predominantly ($>85\%$) via the SMT 
channel (see, e.g., \citealt{Gallegos-Garcia2021,Olejak2021a,Marchant2021,Picco2024}).
We designate the latter type of population as `SMT-dominated'.

\subsection{StarTrack models}

For a second model of the population of He+BH systems (later BBH mergers), 
we use the StarTrack rapid population synthesis code \citep{Belczynski2008,
Belczynski2020}, in particular, the default model described in \cite{Olejak2024}. 
This model incorporates updated mass transfer stability criteria \citep{
Pavlovskii2017,Olejak2021a}, leading to BBH mergers predominantly formed 
via the stable mass transfer (SMT) scenario.

During mass transfer onto a non-degenerate accretor, we assume a fixed 
accretion efficiency of $\beta = 50\%$, where $\beta$ represents the 
fraction of transferred mass that is accreted by the companion star 
\citep{Vinciguerra2020}. Any non-accreted mass is expelled from the 
system, carrying away the specific angular momentum of the binary 
\citep{podsiadlowski1992}. For mass transfer onto a BH, we adopt the 
analytical approximations provided by \citet{King2001}, as implemented 
by \cite{Mondal2020}. This results in highly nonconservative mass transfer, 
where the accretion rate onto the BH is capped at the Eddington limit 
\citep[see, e.g.,][]{King2023}. In this scenario, the non-accreted mass 
is lost with the specific angular momentum of the accretor.

We employ the rapid supernova (SN) engine model with a mixing parameter 
$f_{\rm mix} = 2.5$, consistent with the convection-enhanced SN engines 
proposed by \cite{Fryer2022}. BH formation in our model is accompanied 
by natal kicks drawn from a Maxwellian velocity distribution with a 
dispersion $\sigma = 265 $kms$^{-1}$\citep{Hobbs2005}. However, the kick 
magnitudes are reduced by the amount of fallback, following the prescriptions 
of \cite{Fryer2012} and \cite{Belczynski2012}, which makes significant 
natal kicks for massive BHs unlikely.

To account for observed massive BH-BH mergers (with $M_{\rm BH}\geq$
50\,$M_{\odot}$), we adopt a high threshold for pair-instability supernovae 
(PSN), assuming that stars with helium core masses exceeding $M_{\rm He} 
> 90 M_{\odot}$ are disrupted \citep{Belczynski2020PSN}. For natal BH spins, 
we assume low but nonzero values in the range $\abhone \approx 0.05-0.15$ 
\citep{Belczynski2020}, based on the assumption of efficient angular momentum 
transport in massive stars \citep{Spruit2002}. Additionally, we include 
efficient tidal spin-up of stripped helium cores in close BH-helium core 
systems with orbital periods shorter than 1.1 days, following the prescriptions 
outlined in \cite{Belczynski2020}. To estimate the natal spin of the 
second-born BH, we use Eq.\,(15) of \citet{Belczynski2022}. 

\subsection{X-ray luminosity}

To determine the X-ray luminosity from the vicinity of the BH, we 
follow the analysis of \citet{Sen2024}, who investigated the X-ray 
emission from BHs with main sequence OB star companions. We adopt 
their assumptions as applicable for hot stripped helium star 
companions, outlined below. 

X-rays can be emitted from an accretion disk around the BH when 
the wind matter from the companion star carries enough angular 
momentum to circularise into an orbit greater than the innermost 
stable circular orbit of the BH \citep{iben1996,Frank2002,Sen2021}. 
The X-ray luminosity $L_{\rm X}$ from the accretion disk is 
estimated by Eq.\,(21) of \citet{Sen2024}, given by 
\begin{equation}
    L_{\rm X} = G \frac{M_{\rm BH} \dot{M}_{\rm acc}}{R_{\rm ISCO}}
\end{equation}
where $G$ is the gravitational constant, $M_{\rm BH}$ is the mass 
of the first-formed BH, $\dot{M}_{\rm acc}$ and $R_{\rm ISCO}$ 
are the mass accretion rate onto the BH and the radius of the 
innermost stable circular orbit of the BH given by Eq.\,(22) and 
Eq.\,(29) of \citet{Sen2024} respectively. 

When an accretion disk cannot form, 
X-rays may be emitted from the BH corona via synchrotron emission, 
given by Eq.\,(31) of \citet{Sen2024}, 
\begin{equation}
    L_{\rm X} = \epsilon \dot{M}_{\rm acc} c^2
\end{equation}
where $\epsilon$ is the radiative efficiency. The radiative 
efficiency of X-ray emission without an accretion disk in turn 
depends on the mass accretion rate and the electron heating 
efficiency within the advection-dominated accretion flow 
\citep{Narayan1995,Xie2012}. Typically, the X-ray luminosity 
from an accretion disk is a few orders of magnitude greater 
than in the absence of an accretion disk. 

The wind velocity of the companion plays a crucial role in determining 
the formation of an accretion disk \citep[Eq. (18) of][]{Sen2021}, 
with the typical wind speeds from O-type stars being too high to 
form a disk \citep[see also][]{Hirai2021}. The larger the wind speed, 
the more difficult it is to form a disk. It has been recently shown 
\citep{Sander2023} that the terminal wind velocity from stripped 
hydrogen-free stars does not follow the simple scaling relation of 
the well-known CAK theory \citep{Castor1975}. Hence, to make a 
conservative estimate of the number of He+BH binaries that can form 
an accretion disk, we take an upper limit to the terminal wind 
velocities for our stripped helium stars from the empirically 
determined terminal wind velocities of Galactic WC stars analysed by 
\citet{Sander2012}. 

For models having stripped stars with a radius of less or more than 
1\,$R_{\odot}$, we take the terminal wind velocity as 5000\,km\,s$^{-1}$ 
or 3000\,km\,s$^{-1}$ respectively (Table\,4 of \citealp{Sander2012}). 
Our assumption ensures that an accretion disk forms in the rarest of 
cases. As such, the predicted X-ray luminosities should be lower 
limits. To derive the X-ray luminosity from synchrotron emission, we 
take the electron heating efficiency to be 0.1\% \citep{Xie2012,Sen2024}. 
This ensures a lower limit on the X-ray luminosity in the absence of 
an accretion disk. For higher electron heating efficiencies (10-50\%), 
the radiative efficiency and X-ray luminosity can be up to two orders 
of magnitude higher \citep{Xie2012,Sen2024}. 

%--------------------------------------------------------------------

\begin{figure*}
    \centering
    \includegraphics[height=6cm,width=0.49\linewidth]{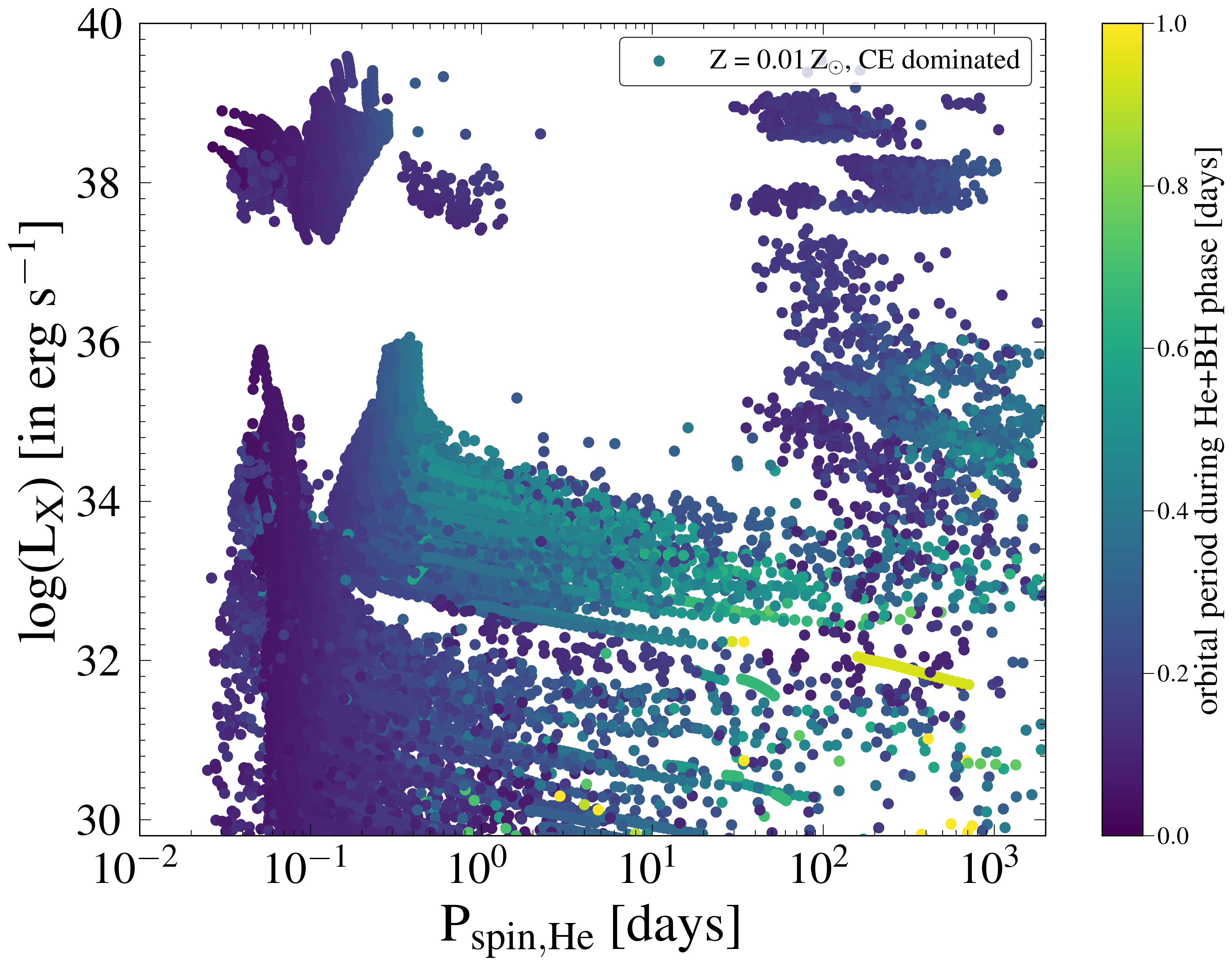}
    \includegraphics[height=6cm,width=0.49\linewidth]{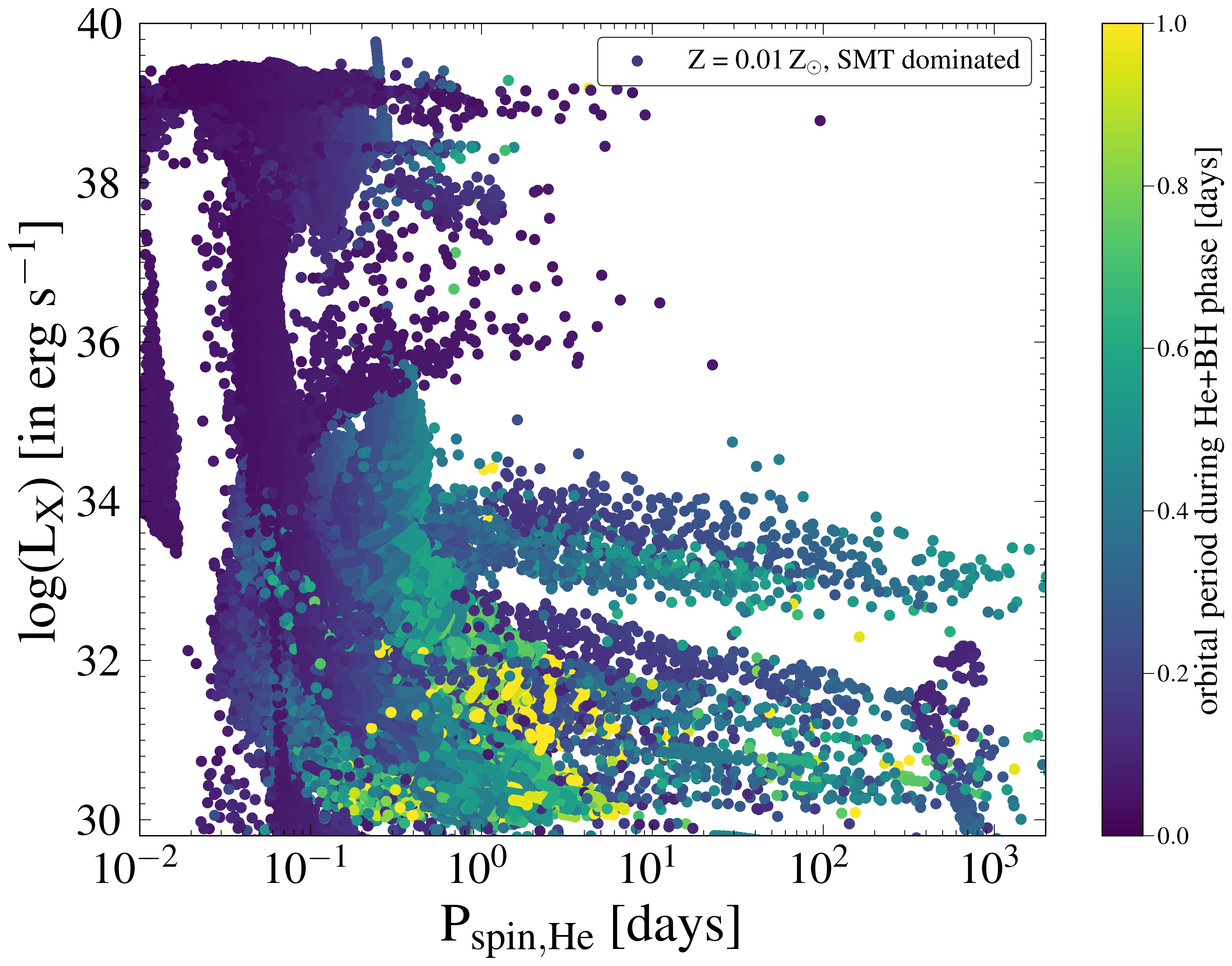}
    \includegraphics[height=6cm,width=0.49\linewidth]{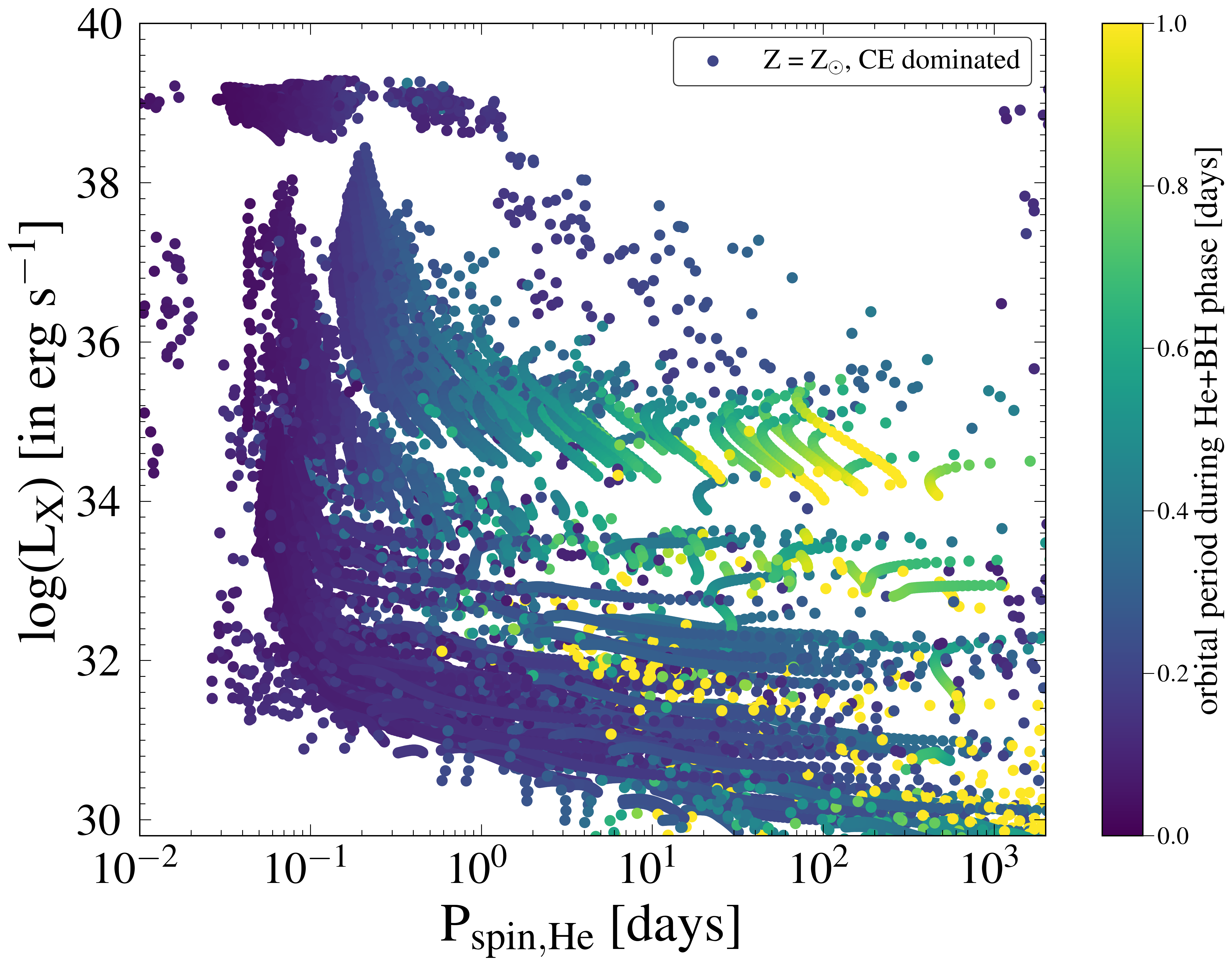}
    \includegraphics[height=6cm,width=0.49\linewidth]{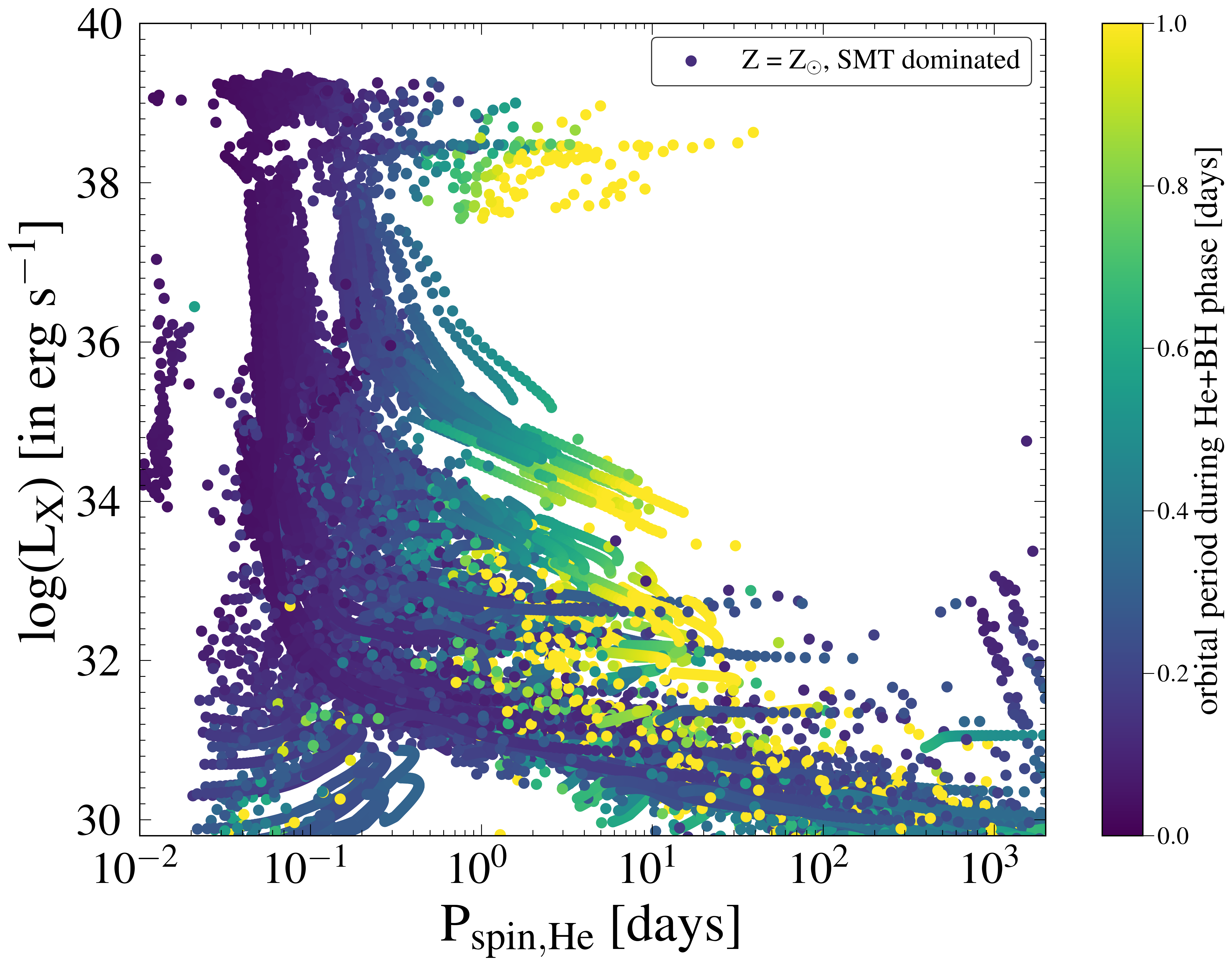}
    \caption{\textbf{BSE models:} Distribution of the estimated X-ray luminosity L$_{\rm X}$ with the spin period of the stripped He star during the He+BH phase (sampled every 0.02\,Myr), for metallicities Z = 0.01\,Z$_{\odot}$ and Z$_{\odot}$ and two criteria of stability of mass transfer (see text). \textit{Colorbar}: all He+BH systems with orbital periods greater than 1\,d that merge within a Hubble time end up in yellow. }
    \label{fig:spinWR_Lx}
\end{figure*}

\section{Results}
\label{section_result}

Here, we investigate the correlation between I) the spin-up of the He 
star during the He+BH phase and II) the natal spin of the second-born 
BH with the X-ray luminosity from the vicinity of the first-born BH 
during the He+BH phase. Our analysis only focuses on those He+BH 
systems that form BH-BH and NS-BH mergers within a Hubble time.

\begin{figure*}
    \centering
    \includegraphics[height=6cm,width=0.49\linewidth]{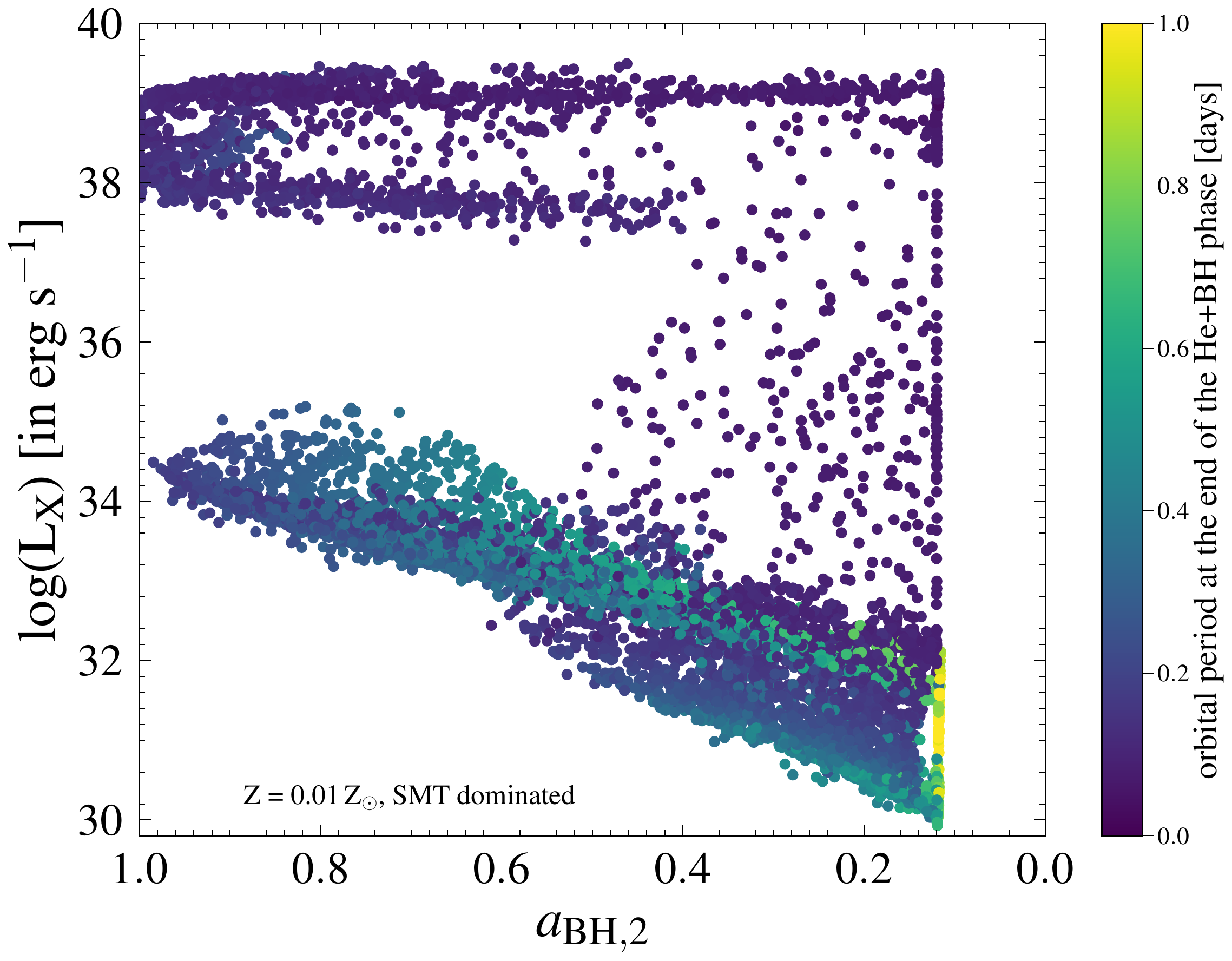}
    \includegraphics[height=6cm,width=0.49\linewidth]{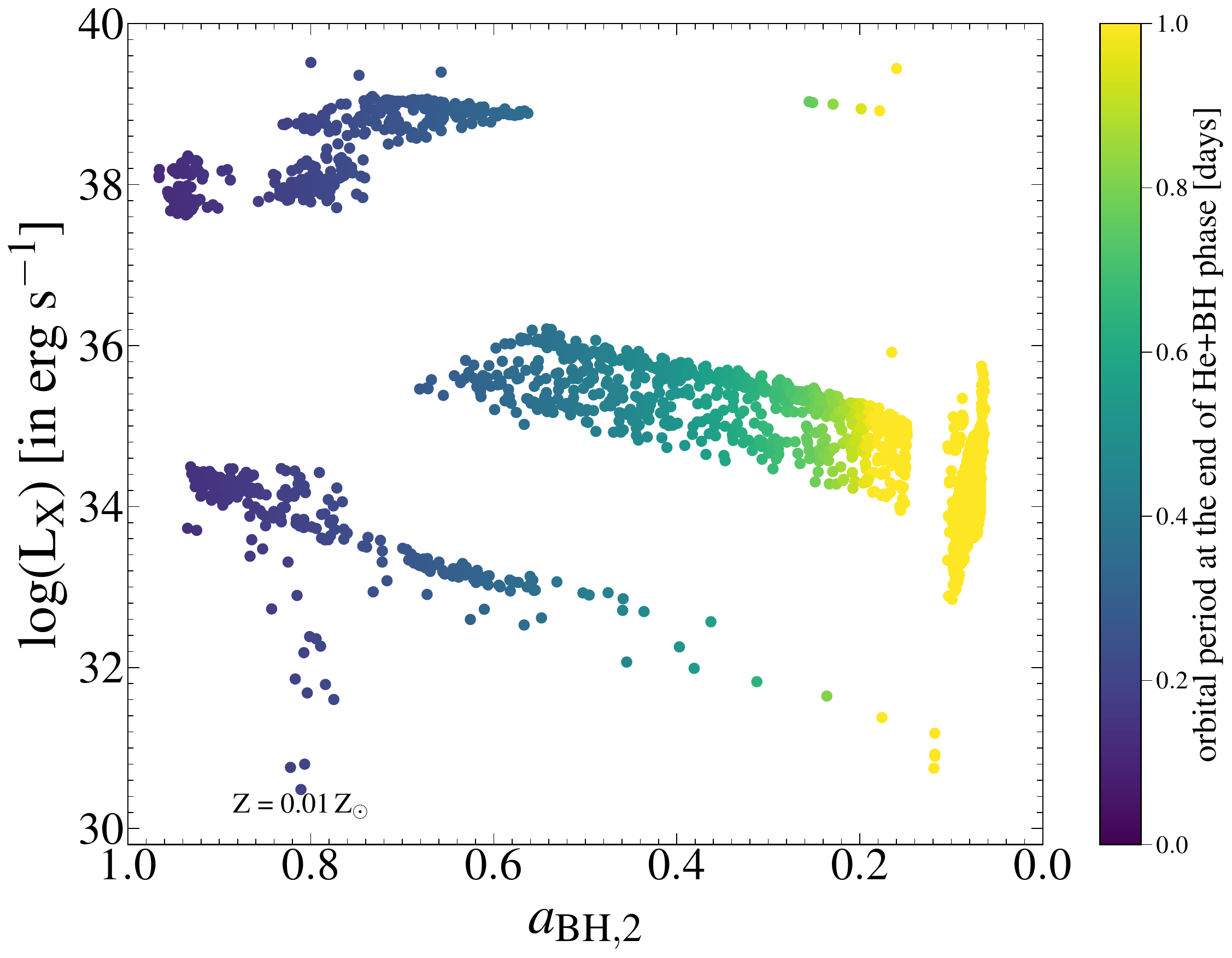}
    \includegraphics[height=6cm,width=0.49\linewidth]{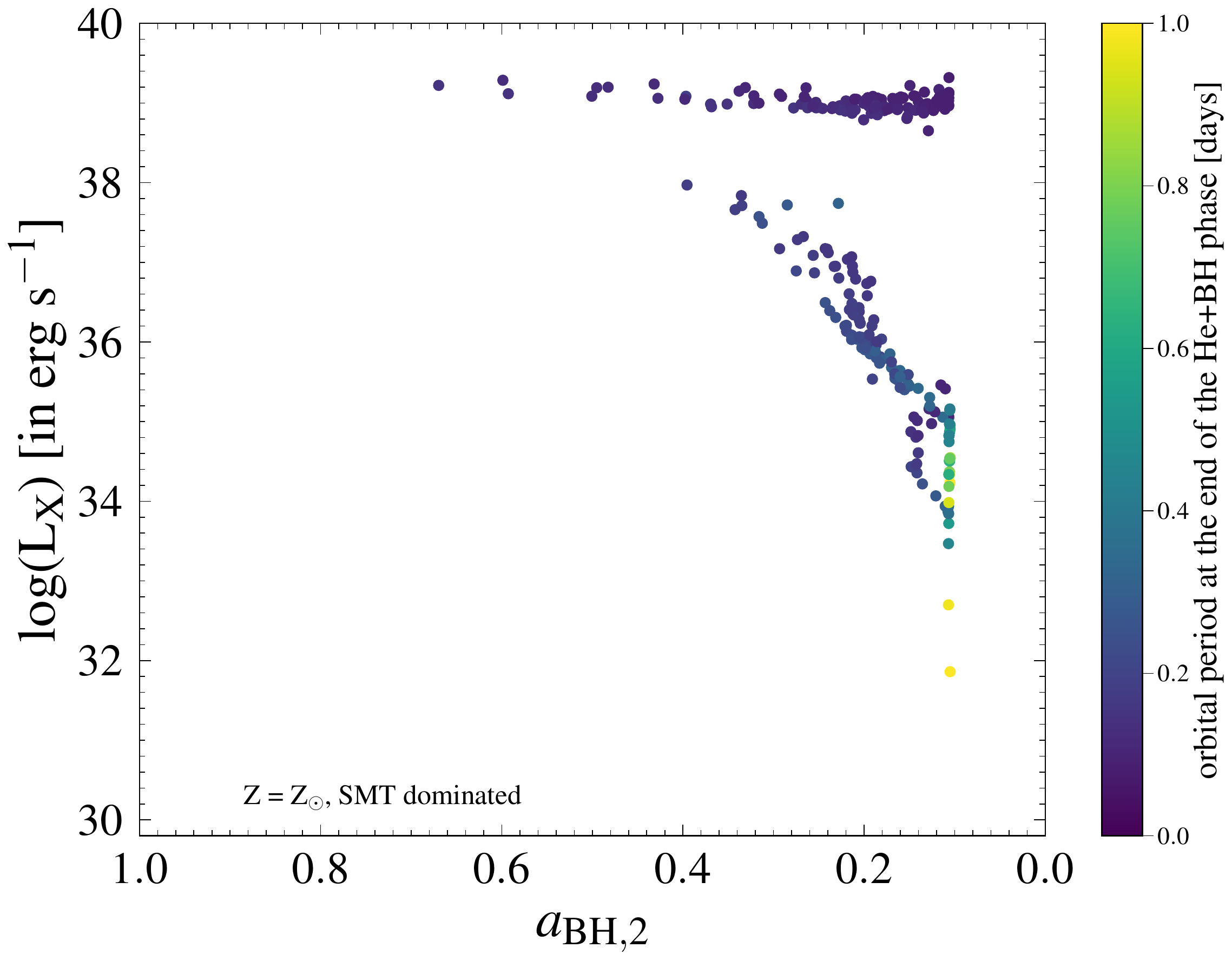}
    \includegraphics[height=6cm,width=0.49\linewidth]{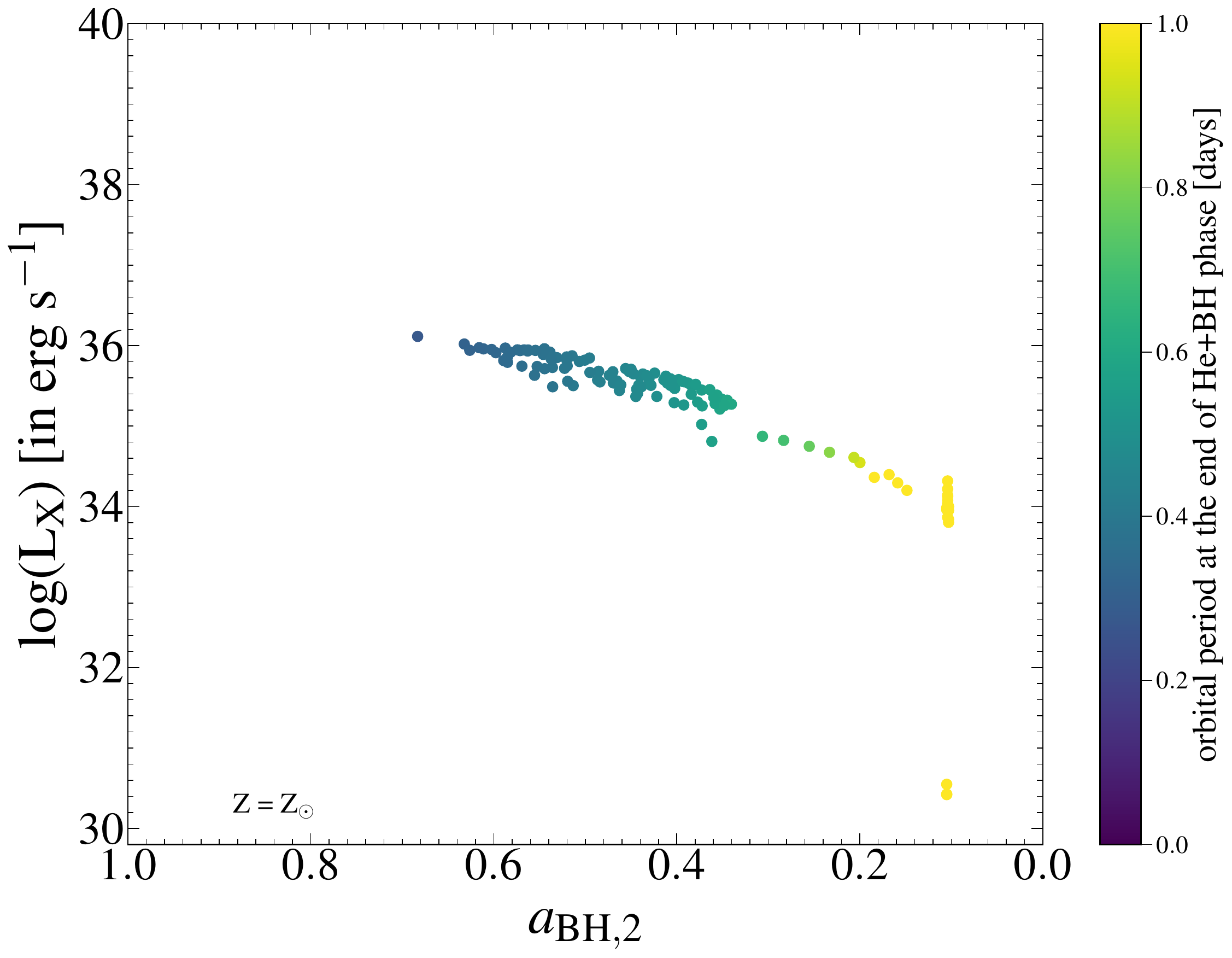}
    \caption{Distribution of the estimated X-ray luminosity L$_{\rm X}$ with the predicted spin of the second-born BH $a_{\rm BH,2}$ at the end of the He+BH phase from BSE (left panels) and StarTrack (right panels), for metallicities Z = 0.01\,Z$_{\odot}$ and Z$_{\odot}$ (top to bottom panels) and the `SMT dominated' case. }
    \label{fig:spinBH_Lx}
\end{figure*}

\subsection{Evolution during the Helium core+black hole phase}

Here, we investigate the X-ray properties of the He+BH binaries 
that will produce a BH-BH or BH-NS merger within a Hubble time. 
Fig.\,\ref{fig:spinWR_Lx} shows the evolution of the spin period of 
the He star with the estimated X-ray luminosity of the firstborn 
BH during the He+BH phase for different sets of binary evolutionary 
models. In all four panels, there is a population of fast-spinning 
($P_{\rm spin}$ < 1\,d) He stars with a BH companion that is expected 
to have X-ray luminosities above 10$^{35}$\,erg\,s$^{-1}$. The shortest 
period He+BH binaries are the most likely to experience strong enough 
tides to spin up the He star \citep{Ma2023} 
and simultaneously favour the formation of an accretion disk from 
which copious amounts of X-rays can be emitted \citep{Sen2021}. 
X-ray luminosities above $10^{37}$\,erg\,s$^{-1}$ are found to arise 
from He+BH binaries where an accretion disk can form around the BH. 

At low metallicity (here 0.01\,Z$_{\odot}$, see also Fig.\,\ref{fig:spinWR_Lx_appendix}), 
for the `CE dominated' set of models (left panel), there is also a 
population of models that have spin periods $\geq1000$\,d and X-ray 
luminosities above $10^{37}$\,erg\,s$^{-1}$. This population, absent 
in the `SMT dominated' channel, %\ola{This luminous subpopulation in 
%the CE dominant model (the upper-right corner of Fig. \ref{fig:spinWR_Lx}), 
%characterized by low spins and tight orbits, 
arises from the fact that the orbital separation at the onset of the 
CE phase was reduced in one timestep to a very short orbital period 
($< 1$\,d). The CE phase is expected to proceed on a dynamical timescale 
with a duration $\leq$ few 1000 yrs (\citealp{Ivanova2013}, for more 
recent works, see \citealp{Hirai2022,DiStefano2023,Gagnier2023,Wei2024,
Vetter2024,Nie2025}). The CE phase is too short for tides to increase the spin velocities
of the He cores significantly. In such cases, stars temporarily retain 
the spin associated with their pre-CE orbit. Once a binary evolves through 
the He+BH phase, the He star can be tidally spun up to much higher values. 
While the CE phase dramatically shrinks the orbital period (from $10^3$ 
days to about 1 day), BBH mergers progenitors in the SMT channel 
typically reduce separation during RLOF (specifically, the one from 
which the He+BH system emerges) less significantly and more 
gradually. 

We find that at Z = 0.01\,Z$_{\odot}$, 0.1\,Z$_{\odot}$, 
0.5\,Z$_{\odot}$ and Z$_{\odot}$, the lifetime-weighted 
fraction of He+BH binaries that have X-ray luminosities higher than 
$10^{35}$\,erg\,s$^{-1}$ is 0.457, 0.412, 0.506 and 0.202 for the `CE 
dominated' channel and 0.346, 0.410, 0.298 and 0.147 for the `SMT 
dominated' channel, respectively. This implies that a large fraction 
of short-period (< 1\,d, \citealp[e.g.][]{Ma2023}) He+BH binaries that 
are expected to spin up due to efficient tides, should also show bright, 
observable X-ray emission at the four investigated metallicities. %Our 
%finding can be used as an observational constraint to study 
%the efficiency of tidal forces to spin up the progenitor of the second 
%formed BH and reproduce the tail of the BH spin distribution via the 
%isolated binary evolution channel. 

Few Wolf-Rayet stars have been identified as non-coronal X-ray emitters 
\citep{Freund2024} in the recently released e-Rosita catalogue \citep{Merloni2024}. 
The X-rays in such systems (e.g. HD 50896, HD 92740, HD 113904) 
may be produced from the accretion of the stellar wind of the 
Wolf-Rayet star on to a compact companion. As such, these are 
interesting targets for follow-up multi-epoch spectroscopy. 
These objects may also be relevant in the context of long-duration 
gamma-ray burst progenitors via the collapsar scenario \citep{Blandford1977,MacFadyen1999,yoon2005,Fryer2022,Gottlieb2024}. 

\subsection{Spin of the second-formed black hole}

Here, we study the correlation between the X-ray luminosity from He+BH 
binaries that will produce a BH-BH merger with the birth spin of the 
second-formed BH.
Fig.\,\ref{fig:spinBH_Lx} shows the predicted X-ray luminosity at the 
last timestep of the He+BH phase and the natal spin of the second 
formed BH. We show the results for two separate rapid binary population 
synthesis codes, BSE (left panels) and StarTrack (right panels). This 
is to improve the robustness of our conclusions to uncertainties in 
stellar and binary evolution physics and its implementation \citep{
Belczynski2022,Agrawal2022}, while acknowledging that accounting for 
the differences are beyond the scope of this work. 

\begin{figure*}
\centering
\includegraphics[width=0.48\linewidth, draft=False]{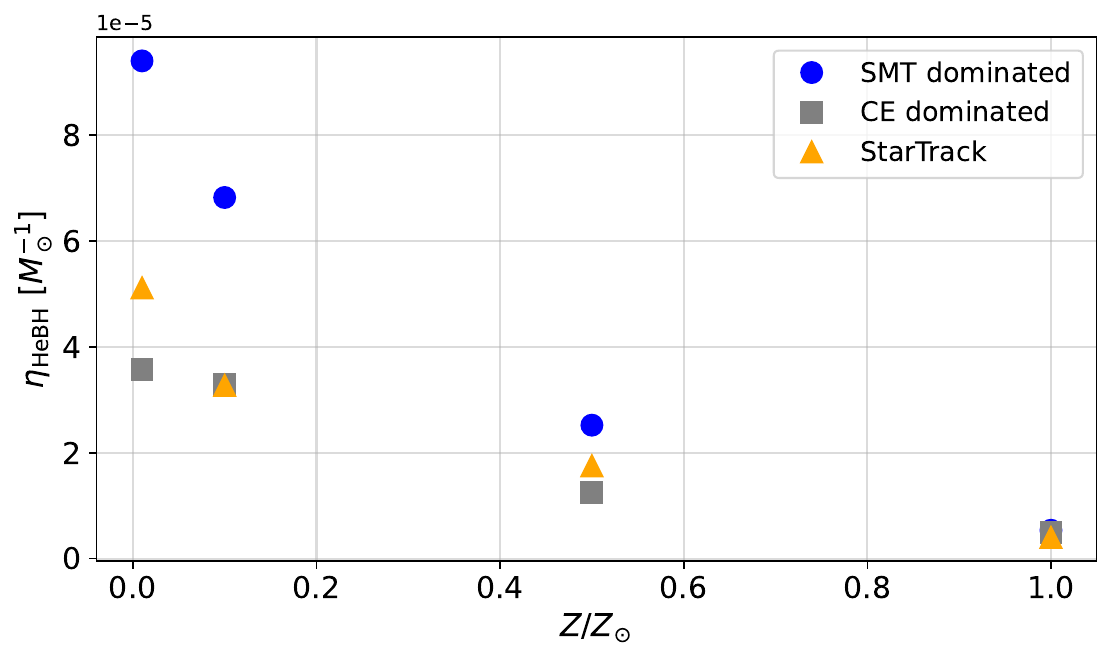}
\includegraphics[width=0.48\linewidth, draft=False]{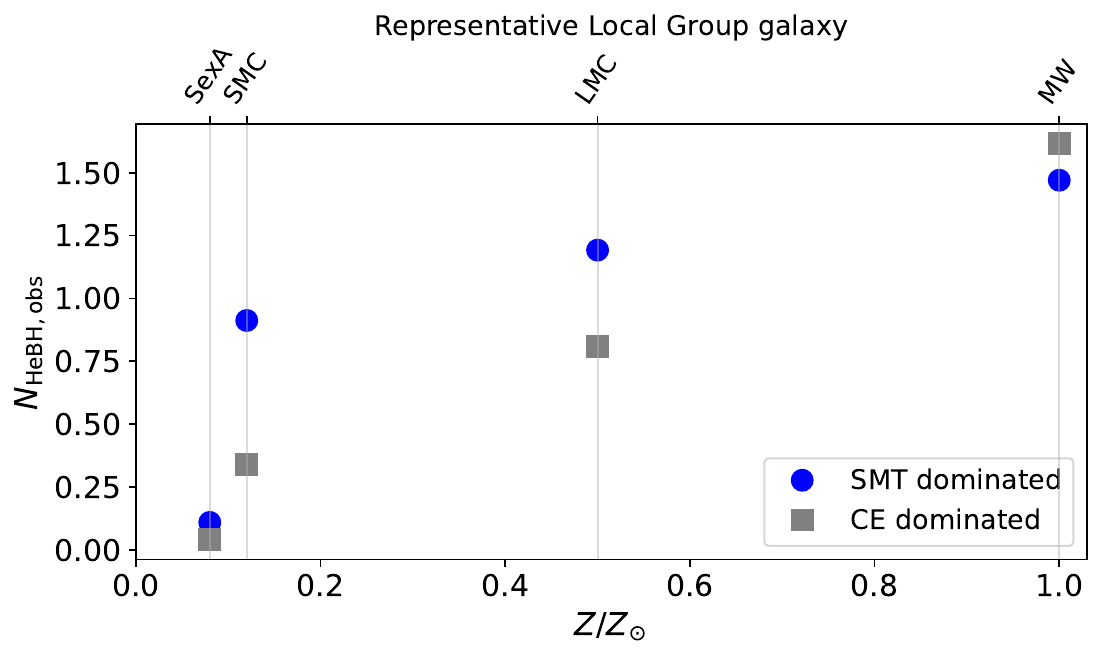}
	\caption{{\bf Left panel:} formation efficiency, $\etahe$, %, and galaxy-wide intrinsic count,
%$\nhe$ (right panel), 
of He+BH binaries as a function of metallicity, $Z$, as obtained from the different
binary-evolutionary models computed in this study (legend). Note that, here, $\etahe$
comprises only the population of He+BH binaries that evolve into BBH or NS+BH mergers within a Hubble time. 
	{\bf Right panel:} Estimated galaxy-wide counts, $\nheobs$, of bright, observable X-ray sources
(with $L_X>10^{35}{\rm~erg}{\rm s}^{-1}$) corresponding to the Local Group galaxies (and their individual SFRs, see text) named along
the upper X-axis. Number estimates from StarTrack are not shown,
as evolutionary data for the first and last time steps of the He+BH phase are only available.}
\label{fig:rates}
\end{figure*}

The contribution of the classical CE channel to the population of BBH 
merger progenitors has been challenged by recent literature after revisiting 
mass transfer stability during Roche-lobe Overflow \citep{Pavlovskii2017,
Gallegos-Garcia2021,Shao2021,Ge2023} and the expected strict conditions 
for successful CE ejection \citep{Klencki2020,Romagnolo2024}. Moreover, 
some studies indicate the SMT channel can reproduce the distribution of 
BH spins of GW population \citep{Olejak2021b} as well the reported 
anti-correlation between $\chi_{\rm eff}$ and mass ratio of GW sources 
\citep{Callister2021,Olejak2024,Banerjee2024}. Therefore, we focus on the 
results for the `SMT dominated' channel, which adopts revised conditions for 
stability of mass transfer, limiting the contribution of the CE channel.

For both BSE and StarTrack, there is a population of rapidly spinning 
second-born BHs ($a_{\rm BH,2}$ > 0.6) with $L_{\rm X} > 10^{37}$\,erg\,s$^{-1}$, 
originating from He+BH binaries that had an accretion disk around the 
first-born BH. These He+BH systems, where an accretion disk can form, show no 
correlation between their X-ray luminosity and the spin of the second-born 
BH. At low X-ray luminosities (see also, Fig.\,\ref{fig:spinBH_Lx_appendix}), 
we find a correlation between the natal spin of the second-born BH and 
the X-ray luminosity from the first-born BH at the end of the He+BH 
phase over $10^{30}$\,erg\,s$^{-1}$$ < L_{\rm X} < 10^{36}$\,erg\,s$^{-1}$, 
although the exact nature of the correlation is unique for each of 
the two rapid population synthesis codes and metallicity. 
The StarTrack models produce much fewer He+BH binaries where an 
accretion disk can form with $L_{\rm X} > 10^{37}$\,erg\,s$^{-1}$. 
However, the mass accretion rates are high enough to produce bright 
X-ray emission from advection-dominated accretion flows near the BH 
corona without an optically thick, geometrically thin disk. 
%On the other hand, the He+BH systems where an accretion can form 
%($L_{\rm X} > 10^{37}$\,erg\,s$^{-1}$) show no correlation between 
%their X-ray luminosity and the spin of the second-born BH, i.e., disk 
%X-ray sources are rather agnostic of $a_{\rm BH,2}$ (or the efficiency of 
%tidal spin-up of the He companion). However, for the dimmer, advective 
%sources, $L_{\rm X}$ is positively correlated with $a_{\rm BH,2}$. 
%In this case, the X-ray luminosity can be used as a measure of the 
%tidal spin-up efficiency. This is true for both BSE and StarTrack 
%models at all metallicities considered in this work. %(Arguably, StarTrack produces much fewer L_X > 10^36 sources.)

For both codes, there is a significant decrease in the number of BBH 
progenitors at Z$_{\odot}$ compared to 0.01\,Z$_{\odot}$. Strong 
stellar winds remove most of the mass of initially massive stars and 
tend to widen the systems such that a BBH merger cannot occur within a 
Hubble time. There is, however, a difference between the subpopulations 
from BSE and StarTrack at high metallicity. In particular, 
the BSE population is dominated by low-spinning BBH merger progenitors ($\abhtwo\leq0.4$), 
while the StarTrack population amply reaches higher BH spins ($\abhtwo\leq0.6$). BSE and 
StarTrack adopts different prescriptions for the final spin of tidally 
locked He stars. In particular, BSE adopts \cite{Bavera_2021b} that 
introduces a dependence on the He star mass, whereas StarTrack uses 
a simpler, mass-independent prescription described in \cite{Belczynski2020}. 
As a result, the \cite{Bavera_2021b} prescription limits the possible 
spin-up of low-mass BH progenitors ($M_{\rm He} \lesssim 10\,M_{\odot}$), 
which are dominant in the high metallicity environment.

Moreover, BSE derives 
the spin of second-born BH by considering the widening of the binary 
orbit during the He+BH phase. As a result, tidally locked systems 
can leave tidal locking during the He+BH phase and end up with low 
natal BH spins. The simplified StarTrack approach derives the second 
born BH spin based on the initial orbital period of the He+BH phase, 
neglecting the eventual widening of the orbit during the He+BH phase. 
A combination of the above two effects leads to contrasting spins of 
predicted populations of BBH mergers at high metallicity for two codes, 
with StarTrack likely overestimating the efficiency of tidal spin-up.

\subsection{Number of observable He+BH systems}
\label{nobs}

The formation efficiency $\etahe$ of He+BH systems that end up as 
binary compact object mergers is given as
\begin{equation}
\etahe\equiv\frac{\nphe}{M_\ast},
\label{eq:etahe}
\end{equation}
where $M_\ast$ is the total simulated stellar mass, after corrections for 
the IMF truncation and binary fraction, corresponding to a particular set 
of evolved binaries. The quantity $\nphe$ is the total number of He+BH systems 
evolving into a double-compact-object merger within a Hubble time that the 
set has produced (for given metallicity and input physical assumptions; 
see Sect.\,\ref{section_method}). For a set of $10^6$ binaries as evolved 
here (Sec.~\ref{section_method}), $M_\ast \approx 9.6\times10^7\Ms$, assuming 
that stars form with a \citet{Kroupa_2001} IMF over the entire ZAMS stellar 
mass range, $0.08\Ms-150.0\Ms$, and with a 100\% binary fraction over the 
simulated ZAMS mass range, $5.0\Ms-150.0\Ms$. Fig.~\ref{fig:rates} (left 
panel) shows the resulting $\etahe$ as a function of metallicity and case, 
for our computed sets.

We assume that the present-day (i.e., at redshift zero) galaxy-wide star 
formation rate (hereafter SFR), $\sfr$, remains constant over the mean lifetime, 
$\tauhe \lesssim 1$ Myr, of such an He+BH phase. The present-day, galaxy-wide 
count of such He+BH binaries is then given by
\begin{equation}
\nhe = \sfr\etahe\tauhe.
\label{eq:nhe}
\end{equation}
%In this study, we adopt a reference SFR of $\sfr=2.0\pm0.7\,\Ms\peryr$, which 
%is the present-day, integrated SFR of the Milky Way galaxy as inferred from 
%infrared observations \citep{Elia_2022}. The resulting $\nhe$ values are 
%$\sim$10s for the Milky Way and the LMC, to $\sim$80-100 for a Milky Way-like 
%galaxy at Z = 0.1\,Z$_{\odot}$ and 0.01\,Z$_{\odot}$. 
%
%are shown 
%in Fig.~\ref{fig:rates} (upper right panel). In summary, we estimate the 
%present-day, inherent, galaxy-wide count of GR-merger-producing He-BH binaries 
%for a Milky Way-like galaxy to lie within a few to 10s, taking into account 
%uncertainties in the binary evolution physics and SFR.
We estimate the count of bright or observable X-ray sources as $\nheobs=f_X\nhe$,
where $f_X$ is the fractional He+BH lifetime with predicted $L_X>10^{35}{\rm~erg}{\rm s}^{-1}$
as obtained from the X-ray luminosities of our evolutionary He+BH binary models.

We estimate $\nheobs$ for Milky Way (MW) and other widely observed Local 
Group galaxies. We adopt the observationally determined $\sfr$ values of $2.0\,\Ms\peryr$ for the
Milky Way \citep{Elia_2022}, $0.2\,\Ms\peryr$ for the Large Magellanic Cloud (LMC) \citep{Harris_2009}, 
$0.05\,\Ms\peryr$ for the Small Magellanic Cloud (SMC) \citep{Schootemeijer_2021}, and $0.006\,\Ms\peryr$ 
for Sextans\,A \citep{Plummer_1995}. Since our computed binary populations
are at discrete metallicities, we match a galaxy with the population that has 
its metallicity closet to the galaxy's observed metallicity. Based on the observed 
average metallicities of the above-considered galaxies \citep{Davies_2015,Choudhury_2016,
Lorenzo2024}, we match MW, LMC, SMC, and Sextans\,A with the \Zsun, 0.5\Zsun, 
0.1\Zsun, and 0.1\Zsun populations, respectively,
assuming \Zsun$=0.02$ as in $\bse$ and StarTrack.
The resulting $\nheobs$ values for these galaxies are shown in Fig.~\ref{fig:rates} (right panel).
Notably, the trend of $\nheobs$ with $Z$ is opposite to that of $\etahe$ with $Z$. For the chosen set of
galaxies, the decline of $\sfr$ with decreasing $Z$ dominates over the increasing trend of $\etahe$.
As such, this simplistic calculation only considers the galaxies' mean metallicities,
ignoring their internal metallicity dispersions.

We calculate that $\nheobs\sim2$ for the Milky Way and 
$\sim1$ for the LMC, as demonstrated in Fig.~\ref{fig:rates} (right panel). 
There is only one observed High-mass X-ray Binary in the Milky Way harbouring 
a Wolf-Rayet star \citep[Cyg\,X-3,][]{Giacconi1967,Zdziarski2013} and none 
observed in the LMC. Our results may imply that the accretion efficiency 
of BHs in orbit with He stars is much lower than assumed in our work or 
very short-period He+BH systems are rarer than predicted by the current 
binary population synthesis calculations. If the latter is true, explaining 
the high-spin tail of GW mergers by invoking the tidal spin-up during the 
He+BH phase may be unlikely. 

Although our predicted number of observable He+BH binaries in the local 
neighbourhood is low, our results are sensitive to adopted assumptions 
on several uncertain processes e.g., the mass transfer physics \citep{
Gallegos-Garcia2021,Olejak2021a,Sen2022,Willcox2023,Picco2024}, stripped-star winds 
\citep{Detmers2008,Sander2023proceedings,Shenar2024} and core-collapse supernova 
\citep{David2022a,David2022b,Fryer2022,Janka2024}. There are also 
significant limitations in predicting detailed properties of the structure 
and radius of helium cores using rapid codes \citep[e.g.][]{laplace2020,
Laplace2021,Wang2024,Hovis-Afflerbach2024}, that may affect our results. 

\section{Conclusions}
\label{section_conclusion}

In the isolated binary evolution channel, the immediate progenitors 
of BBH binaries that can merge within Hubble time are expected to be 
short-period He+BH systems \citep[see][and references therein]{Korb2024}. 
At sufficiently short periods (< 1\,d), it has been proposed that 
the He star companion to the BH can be tidally spun up such that the 
resultant BH formed from the collapse of the He star can have a high 
Kerr parameter \citep[e.g.][]{Detmers2008,Qin2018,Fuller2022,Ma2023}. 
Such an evolutionary channel has been invoked to explain the high 
effective spin parameters of a few BBH mergers observed by LIGO/Virgo 
\citep{Belczynski2020,Olejak2021b}. However, 
detections of highly spinning He stars in short-period binaries with 
BHs have not been unambiguously confirmed to prove the existence of this 
formation channel. 

In this work, we investigate the possibility of X-ray emission from 
very short-period He+BH binaries as an observational constraint to 
investigate the detectability and formation efficiencies of the 
progenitors of BBH mergers. We use recently published prescriptions 
\citep{Sen2021,Sen2024} to estimate lower limits to the X-ray 
luminosity from the vicinity of the BH in short-period He+BH binaries. 
We post-process the X-ray recipe into two rapid population synthesis 
codes, BSE and StarTrack, to study populations of such binaries at 
four metalicities Z = Z$_{\odot}$, 0.5\,Z$_{\odot}$, 0.1\,Z$_{\odot}$, 
and 0.01Z$_{\odot}$. 

Our population synthesis calculations predict that a significant 
fraction (10-50\%) of GW merger producing He+BH binaries should 
produce X-ray luminosities in 
excess of 10$^{35}$\,erg\,s$^{-1}$ (i.e. observable with current 
X-ray telescopes) at all metallicities, from both 
BSE and StarTrack. The intrinsic formation efficiencies of such 
observable He+BH systems increase with a decrease in metallicity. Because evolutionary 
selection effects favour the formation of BBH merger progenitors 
at low metallicity \citep[see also][]{VanSon2025} and observational selection effects only allow 
the detection of the brightest X-ray sources, our models predict 
$\sim$2 and $\sim$1 He+BH binaries in the Milky Way and the LMC 
to be detectable in X-rays. However, our work only investigates 
the observability in X-rays of He+BH binaries that will produce a 
GW merger within Hubble time. As such, our predictions are lower 
limits since the He+BH binaries that do not evolve to form merging 
GW binaries may still produce observable X-rays during their He+BH 
phase. 

Upcoming X-ray surveys of the LMC \citep{Antoniou2022} and Sextans\,A 
\citep{Antoniou2023}, and the ongoing X-ray surveys of the Milky Way 
that have identified WR stars with untypical coronal emission 
\citep{Freund2024} may provide the required observational leads to 
identifying the population of X-ray-emitting short-period He+BH 
binaries. Combined with follow-up optical studies of Sextans\,A 
\citep{Lorenzo2024}, empirical constraints on the the efficiency 
of the tidal spin-up of He stars during the He+BH phase as a possible 
channel to explain the highly effective spin parameters of some BBH 
mergers may be possible. They could also be interesting systems as 
progenitors of long-duration gamma-ray bursts \citep{Blandford1977,
MacFadyen1999,yoon2005,Fryer2022,Gottlieb2024}.

%--------------------------------------------------------------------

\begin{acknowledgements}
      We thank the anonymous referee, Mathieu Renzo and Norbert Langer for insightful comments that improved the manuscript. KS is funded by the National Science Center (NCN), Poland, under grant number OPUS 2021/41/B/ST9/00757. AO acknowledges funding from the Netherlands Organisation for Scientific Research (NWO), as part of the Vidi research program BinWaves (project number 639.042.728, PI: de Mink). SB acknowledges funding for this work by the Deutsche Forschungsgemeinschaft (DFG, German Research Foundation) through the project ``The dynamics of stellar-mass black holes in dense stellar systems and their role in gravitational wave generation'' (project number 405620641; PI: S. Banerjee). Part of the computations presented here have been performed on the Marvin HPC facility of the University of Bonn. SB acknowledges the generous support and efficient system maintenance of the computing teams at the AIfA, HISKP, and the University of Bonn. 
\end{acknowledgements}

\bibliographystyle{aa}
\bibliography{faint}

\onecolumn

\begin{appendix}

\section{Supplementary plots}

\begin{figure*}[!h]
    \centering
    \includegraphics[height=5.35cm,width=0.46\linewidth]{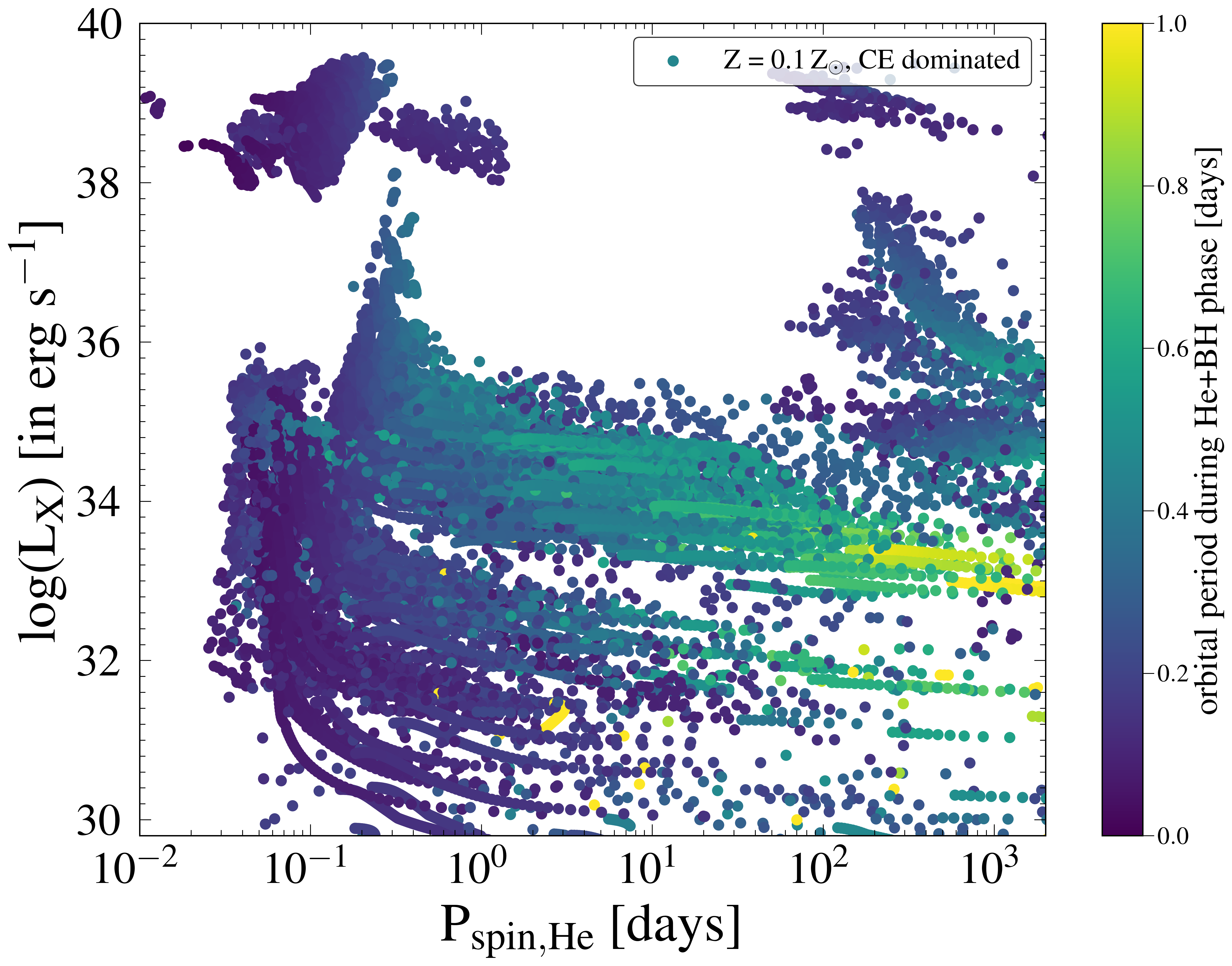}
    \includegraphics[height=5.35cm,width=0.46\linewidth]{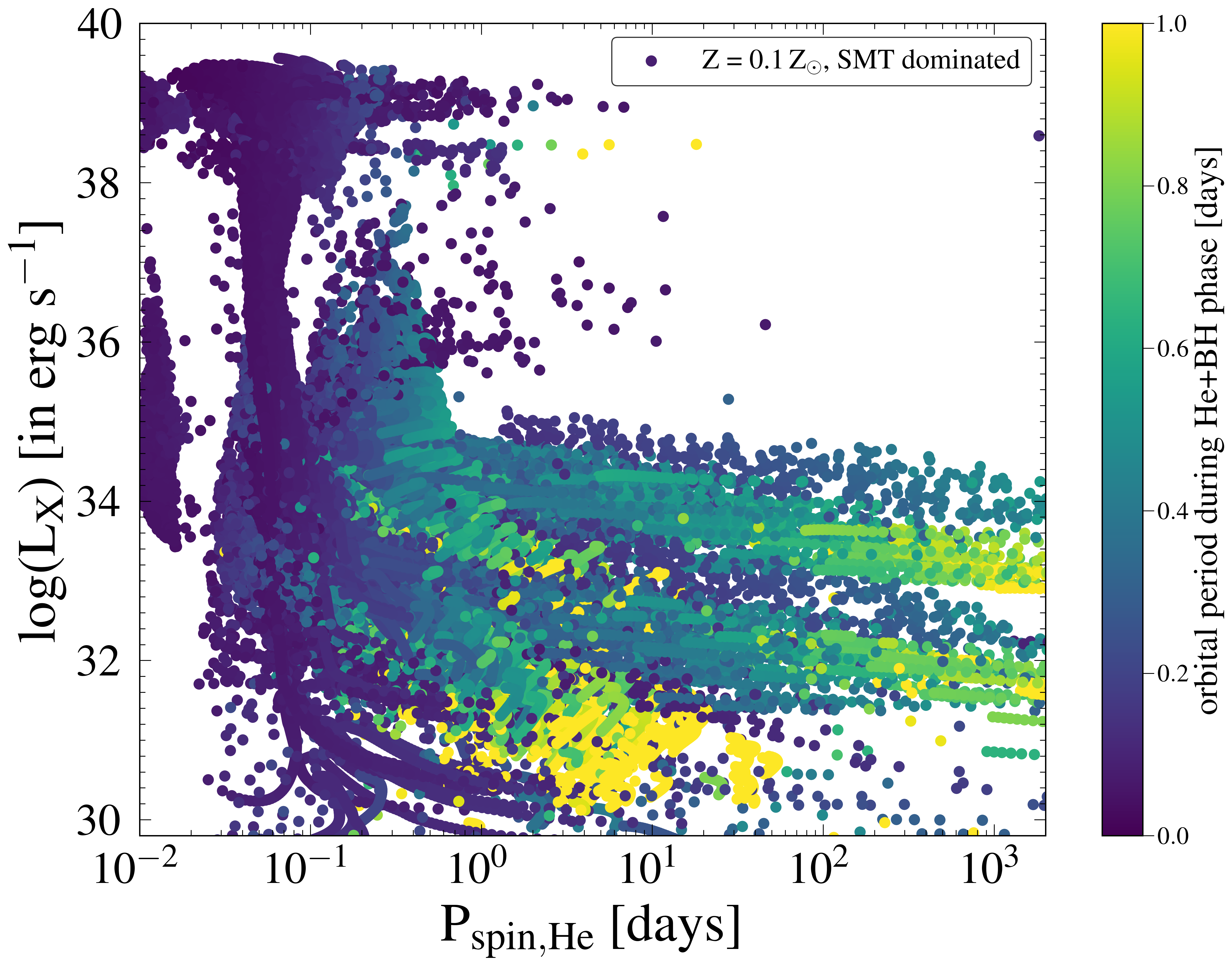}
    \includegraphics[height=5.35cm,width=0.46\linewidth]{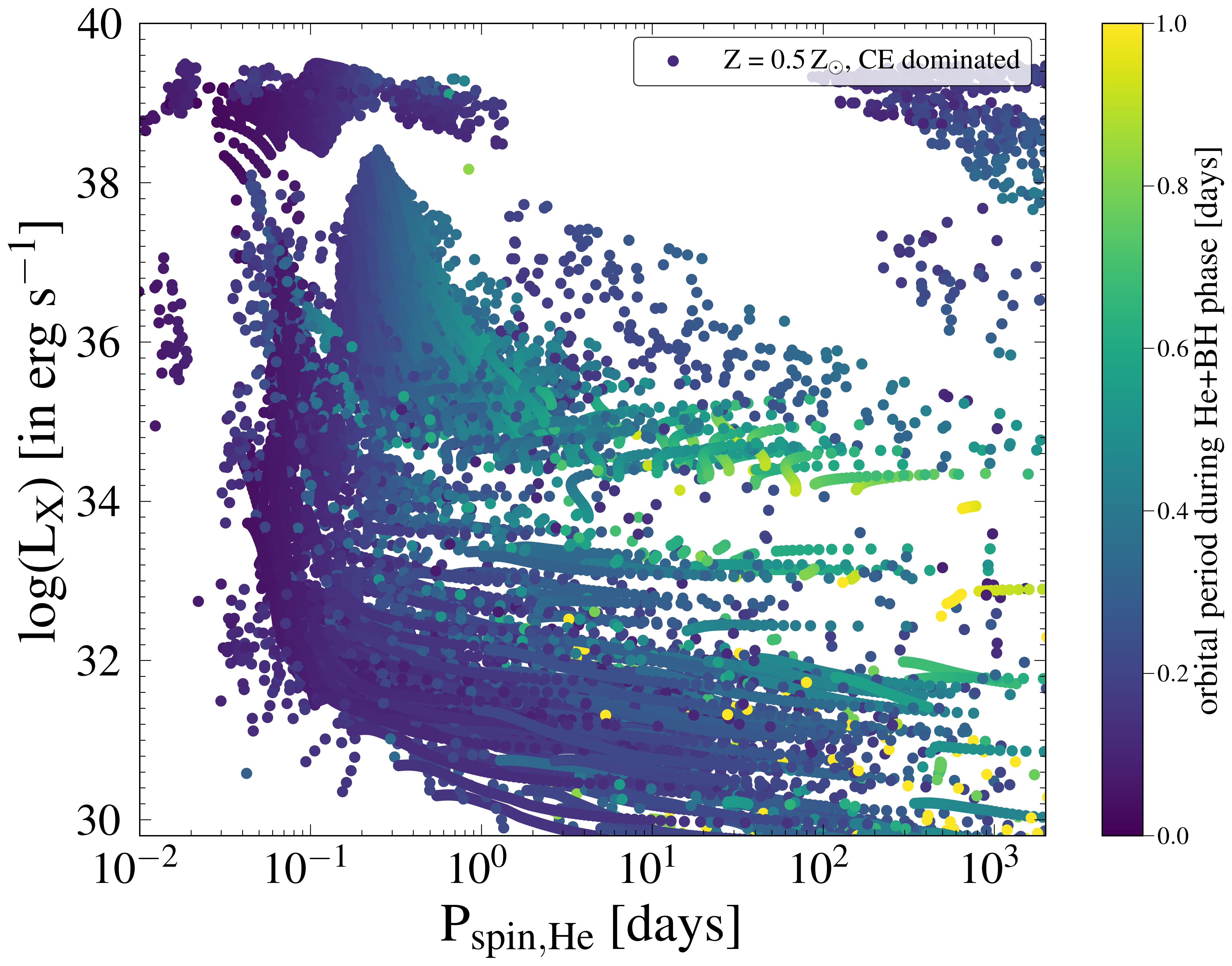}
    \includegraphics[height=5.35cm,width=0.46\linewidth]{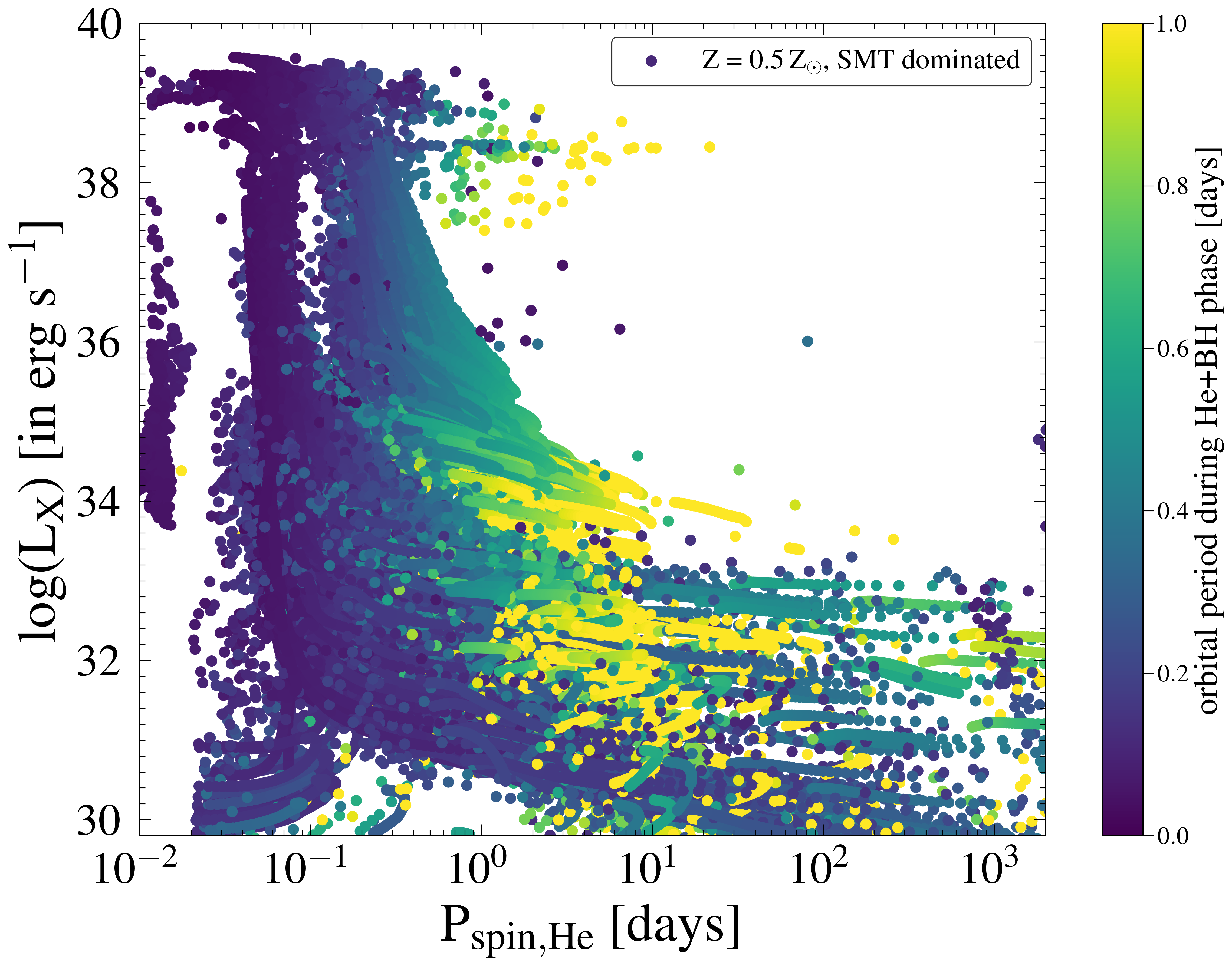}
    \caption{Same as Fig.\,\ref{fig:spinWR_Lx}, but for Z = 0.1\,Z$_{\odot}$ and 0.5\,Z$_{\odot}$. }
    \label{fig:spinWR_Lx_appendix}
\end{figure*}

\begin{figure*}[!h]
    \centering
    \includegraphics[height=5.35cm,width=0.46\linewidth]{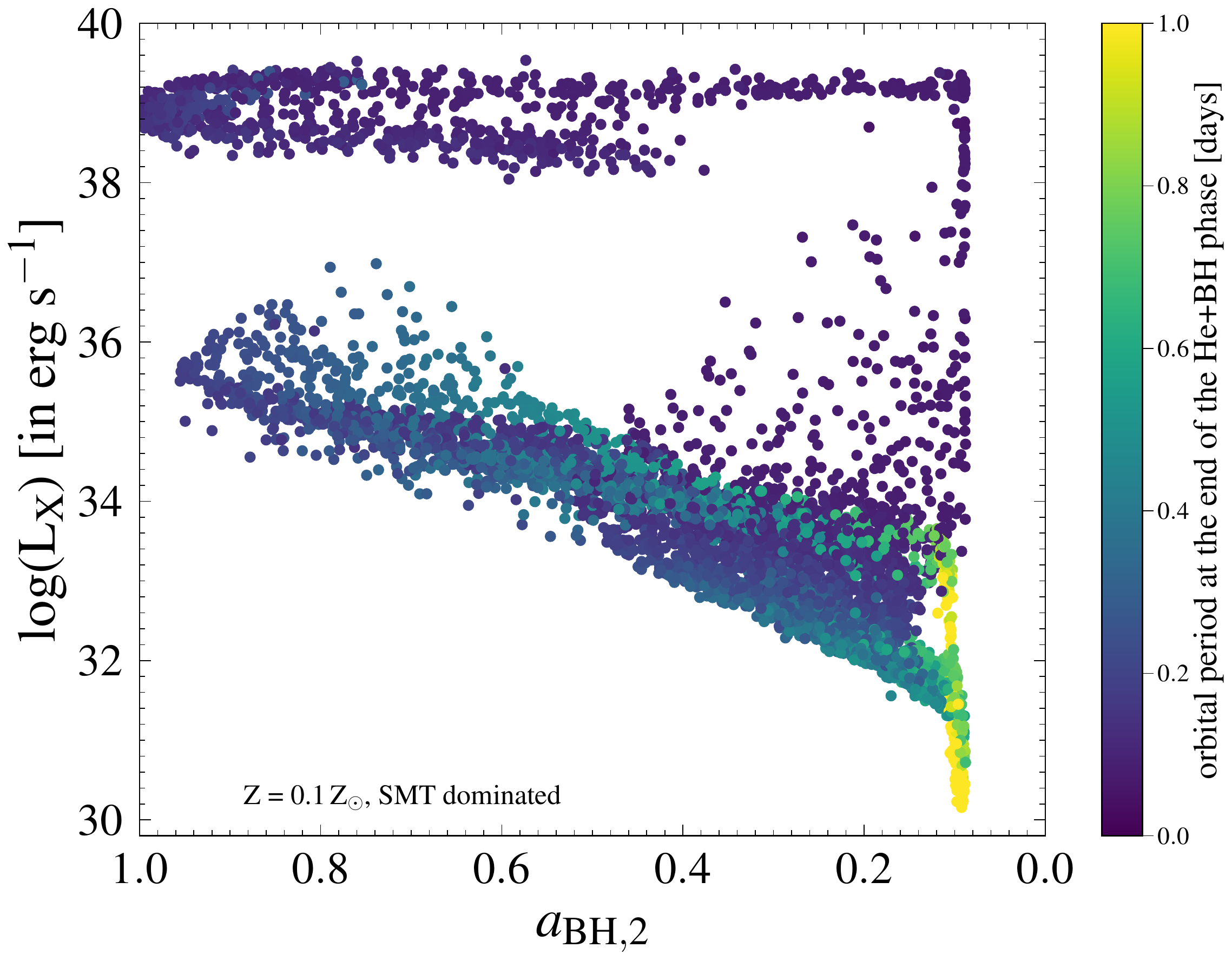}
    \includegraphics[height=5.35cm,width=0.46\linewidth]{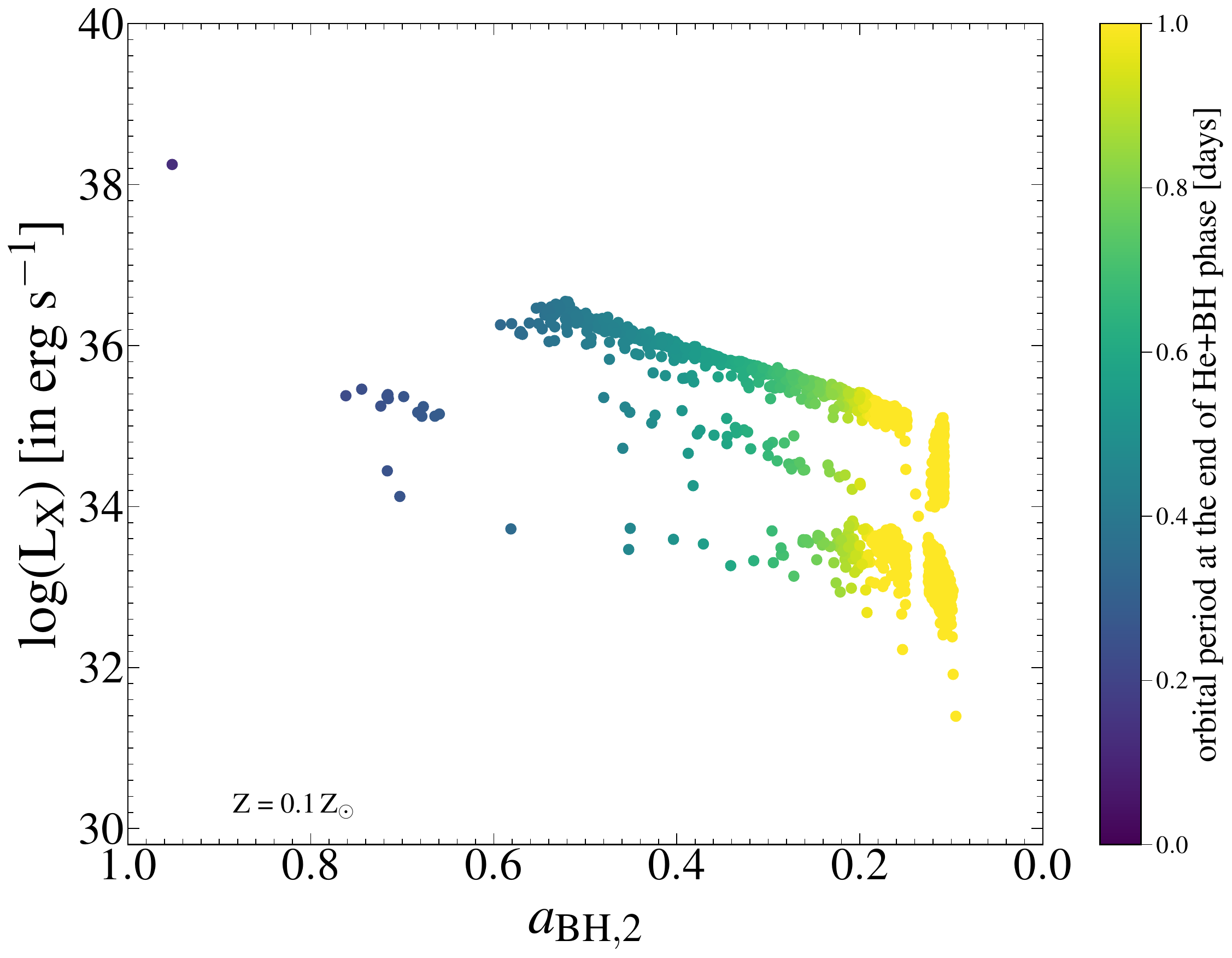}
    \includegraphics[height=5.35cm,width=0.46\linewidth]{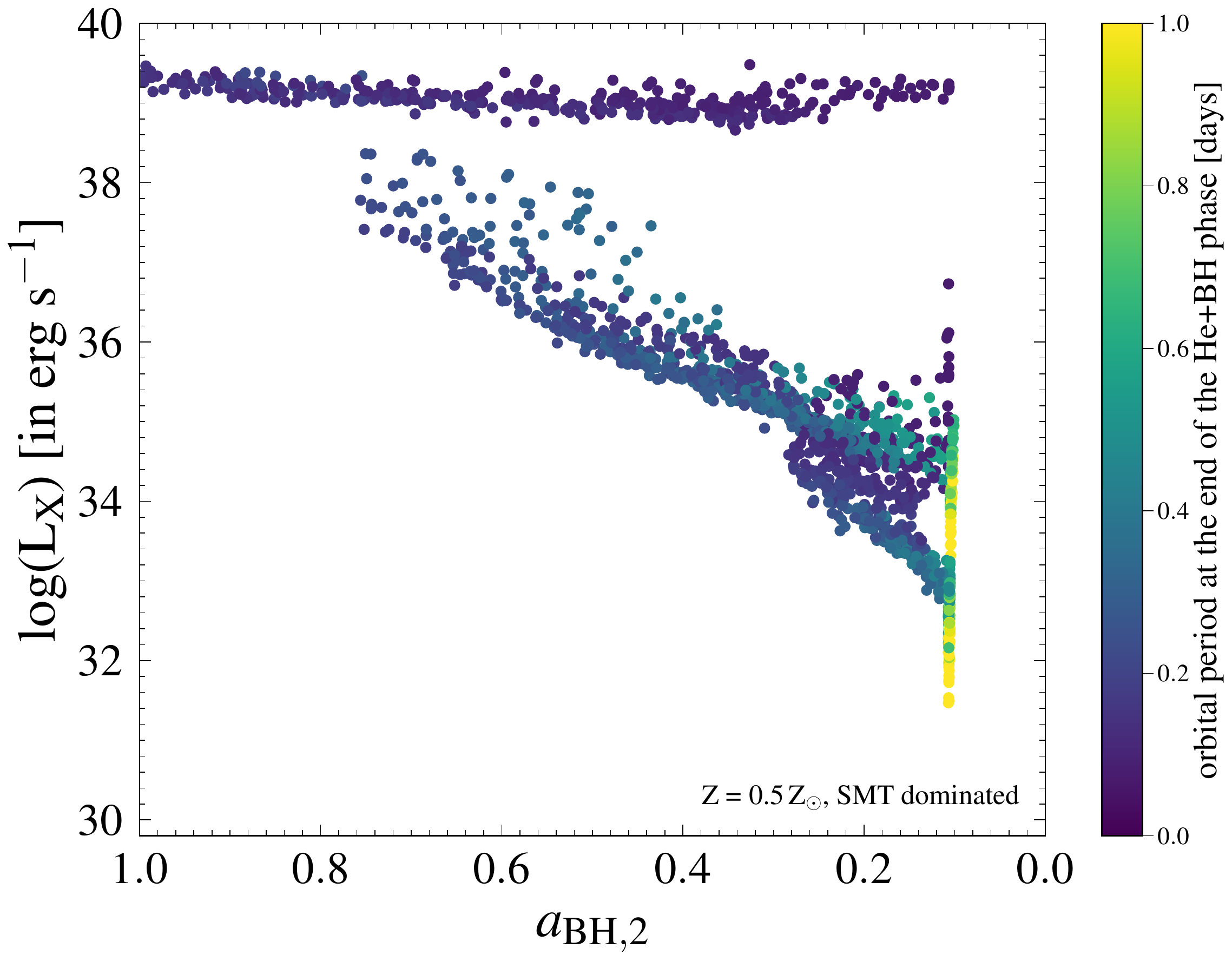}
    \includegraphics[height=5.35cm,width=0.46\linewidth]{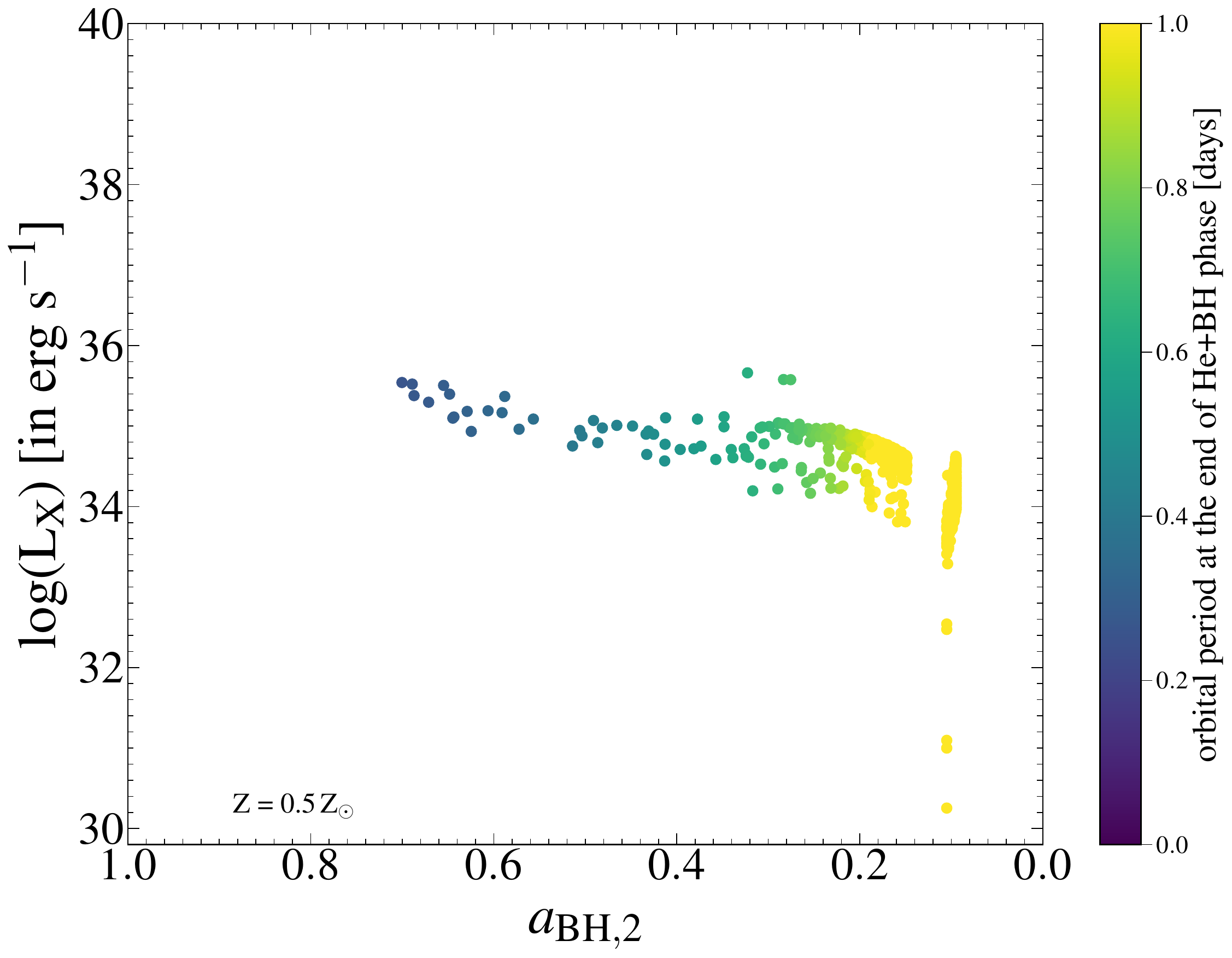}
    \caption{Same as Fig.\,\ref{fig:spinBH_Lx}, but for Z = 0.1\,Z$_{\odot}$ and 0.5\,Z$_{\odot}$. }
    \label{fig:spinBH_Lx_appendix}
\end{figure*}

\end{appendix}

\end{document}